\documentclass[12pt]{article}

\usepackage{amsmath,amsfonts,amssymb,amsthm}
\newcommand{\R}{\mathbb{R}}
\newcommand{\CC}{\mathbb{C}}
\newcommand{\TT}{\mathbb{T}} \newcommand{\ZZ}{\mathbb{Z}}
\newcommand{\NN}{\mathbb{N}}
\newcommand{\p}{\mathbf{p}}

\newcommand{\x}{\mathbf{x}}

\newcommand{\y}{\mathbf{y}}

\newcommand{\kk}{\mathbf{k}}
\newcommand{\T}{\mathbf{t}}
\newcommand{\s}{\mathbf{s}}
\newtheorem{definition}{Definition}[section]
\newtheorem{theorem}{Theorem}[section]
\newtheorem{lemma}{Lemma}[section]

\newtheorem{proposition}{Proposition}[section]

\begin{document}

\title{On the Spectral and Wave  Propagation Properties of the
Surface Maryland Model}
\author{F. Bentosela, Ph. Briet
\\Centre de Physique Th\'{e}orique, Luminy, case 907, Marseille 13288
France,\\L. Pastur \\ University Paris 7, 2, pl. Jussieu, 75251,
Paris, France
%\\ E-mail: \texttt{, bento@cpt.univ-mrs.fr, briet@cpt.univ-mrs.fr,
%pastur@cpt.univ-mrs.fr}
}

\maketitle
\date{}

\begin{abstract}We study the discrete Schr\"odinger operator $H$ in $\ZZ^d$ with
the  surface  potential of the form $V(x)=g
\delta(x_1) \tan \pi(\alpha \cdot x_2+ \omega)$, where for $x \in
\ZZ^d$ we write $x=(x_1,x_2), \quad x_1 \in \ZZ^{d_1}, \ x_2 \in
\mathbb{Z}^{d_2}, \; \alpha \in \R^{d_2}, \; \omega \in [0,1)$.
We first consider the case where the components of
the vector $\alpha$ are rationally independent, i.e. the case of
the quasi periodic potential. We  prove that the spectrum of $H$ on
the interval $[-d,d]$ (coinciding with the spectrum of the discrete
Laplacian) is absolutely continuous.
Then we show that
generalized eigenfunctions corresponding to this interval have the
form of volume (bulk) waves, which are oscillating and non
decreasing (or slow decreasing) in all variables. They are the sum
of the incident plane wave and of an infinite number of reflected
or transmitted plane waves scattered by the "plane" $\ZZ^{d_2}$.
These eigenfunctions are orthogonal, complete and
verify a natural analogue of the Lippmann-Schwinger equation. We
also discuss the case of rational vectors $\alpha$ for
$d_1=d_2=1$, i.e. a periodic surface potential. In this case
we show that the spectrum is absolutely continuous
 and besides  volume (Bloch) waves there are also
surface waves, whose amplitude decays exponentially as $|x_1| \to \infty$.
The part of the spectrum corresponding  to the surface states
consists of a finite number of bands. For large $q$  the bands outside of  $[-d,d]$ are
exponentially small in $q$, and converge in a natural sense to
the pure point spectrum, that was found in \cite{KP} in the case
of the Diophantine $\alpha$'s.

\end{abstract}

\newpage

\section{Introduction }
\setcounter{equation}{0}

The progress of the last decades in spectral theory
of differential and finite difference
operators with random ergodic and almost periodic coefficients in
the whole space makes natural the study of  operators
with same type of coefficients supported on a subspace only.
 Being of evident interest from the point of view of wave physics, they
 provide a class of operators
''intermediate'' between operators whose coefficients decay in all
coordinates (scattering theory) and operators, having coefficients
of the same order of magnitude in all coordinates. We
mention recent papers \cite{BS}, \cite{Gri}, \cite{KP}, \cite{JL1}
- \cite{JM4}, \cite{JMP}, \cite{KP},
devoted to the study of the spectral and
 related properties for operators of such a kind. These operators are either defined on the half-space by
random, almost periodic or periodic boundary conditions or have
the same type of coefficients  supported on  certain subspaces  of
 $\R^{d}$ or  $\ZZ^{d}$.

\medskip \noindent  As in \cite{KP}  we consider here the discrete
 Schr\"odinger
operator
\begin{equation}\label{ham}
H=H_0+V
\end{equation}
acting on $l^2(\ZZ^{d})$, where
\begin{equation}\label{lap}
(H_0\Psi)(\x) = -1/2 \sum_{\vert \x- \y \vert =1} \Psi(\y),
\end{equation}
is the discrete Laplacian,
\begin{equation}\label{Vt}
V(\x)=\delta(x_1)v(x_2),\quad \x=(x_1, x_2),\quad x_1 \in
\ZZ^{d_1},\quad x_2 \in \ZZ^{d_2}, \quad d_1+d_2=d,
\end{equation}
with
\begin{equation}\label{vt}
v(x_2) = g \tan \pi\big(\alpha \cdot x_2+\omega)
\end{equation}
is the multiplication operator ("surface" potential), whose
support is the subspace $\mathbb{Z}^{d_2}$ of the space
$\mathbb{Z}^{d}$, and
\begin{equation}\label{par}
d_1,d_2 \in \mathbb{N}, \; g>0, \; \alpha \in \mathbb {R}^{d_2},
\; \omega \in [0,1)
\end{equation}
are the parameters determining the potential.

\medskip \noindent It was shown in \cite{KP} that for any
$g\not= 0$, $\omega \in
[0,1)$, and for $\alpha \in \R^{d_2}$, satisfying the Diophantine
condition i.e. there exists $\varepsilon >0$ such that
\begin{equation}\label{di}
\vert \alpha \cdot x_2- m\vert \ge \mathrm{const} / \vert
x_2\vert^{d_2+ \varepsilon},  \quad \forall x_2 \in \ZZ^{d_2}
\setminus \{0\}, \quad \forall m \in \ZZ,
\end{equation}
the spectrum of $H=H_0+V$, lying outside the spectrum $[-d,d]$ of
the discrete Laplacian (\ref{lap}), is pure point, dense, of
multiplicity one, and the respective eigenfunctions decay
exponentially at infinity.

\medskip \noindent The  "volume" version of this operator,
corresponding to the case $ d_1=0 $, has been studied earlier in
\cite{PF:84,Si:84}. The operator has a complete system of
exponentially decaying eigenfunctions, corresponding to the pure
point dense spectrum of multiplicity one occupying the whole real
axis. This spectral structure is caused by strong and irregular
fluctuations of the quasi periodic potential (\ref{Vt}). It is the
extreme case of the strong localization regime, which  in general
appears  either if, for a fixed energy, the amplitude of the
potential (random or almost periodic) is large enough or, if for a
fixed potential, the energy is close enough to the spectrum edges
(see \cite{PF} for related results and references).

\medskip \noindent In the case $d_1=1$ the support of the
potential is the hyperplane $\mathbb{Z}^{d-1}$ of the space
$\mathbb{Z}^{d}$. This is why it is natural to call the respective
operator (\ref{ham}) - (\ref{par}) the surface Maryland model.
This operator is closely related to the boundary value problem
(\ref{H1}), considered in \cite{JMP,JM1,JM4}. We may also call the
operator (\ref{ham}) - (\ref{par}), for $d_1 \ge 2$, the subspace
Maryland model.

\medskip
\noindent These models can be analyzed in great detail,
thereby providing examples of spectral types which are only partly
known for general random or almost periodic function $v$ in
(\ref{Vt}). All these versions of the Maryland model have an
absolutely continuous component of the spectrum. This component was
first indicated in \cite{JMP}, and then was studied in \cite{JM1}
in the context of the boundary value problem defined by (\ref{vt})
and by formula (\ref{H1}) below. It was proven that if the
components of the vector $\alpha \in \R^{d_2}$ are rationally
independent, this part of the spectrum of $H$ is purely
absolutely continuous and also that the properly defined wave operators
corresponding to this part exist and are complete. Besides, it was
proven in \cite{JM4} that the surface states (see \cite{JMP,JM4}
for definitions) are absent.

\medskip \noindent
In this paper we develop several general ideas and
results of the theory by considering the explicitly soluble model,
defined by formulas (\ref{ham}) - (\ref{par}). We begin by showing
that the Green function of the model can be written in a rather
convenient form (Section 2). By using this form we study first the
quasi-periodic case of rationally independent components of the
vector $\alpha $ in (\ref{Vt}) (Section 3). We prove that the
spectrum of the operator is purely absolutely continuous on the
interval $[-d,d]$ (on the spectrum of discrete Laplacian) and that
the wave operators, corresponding to this part of the spectrum
exist (these facts were proved in \cite{JM1,JM4} by other
methods). Then we find an explicit form of the generalized
eigenfunctions (polynomially bounded solutions of
the respective equation), corresponding to this part of the spectrum. These
eigenfunctions possess properties, similar to those of the
Sommerfeld solutions of  scattering theory. Along the $x_2$ direction,
they behave like Bloch-Floquet solutions. They are orthogonal and
complete on the interval $[-d,d]$ of the spectrum.
 As they do not decay in the
longitudinal coordinates $x_{1}$ we call them volume states.
We consider also the case of rationally dependent components of 
the vector
$\alpha $ in (\ref{vt}), where the respective surface potential is
periodic in $x_{2}$, restricting ourselves to the technically
simplest case of $d_{1}=d_{2}=1$ (Section 4). In this case the
whole spectrum is absolutely continuous. It consists of the interval
$[-d,d]$ as in the quasiperiodic case, and of a certain number of intervals,
some of them possibly intersecting $[-d,d]$. To the interval $[-d,d]$ 
correspond
generalized eigenfunctions which do not decay in the
longitudinal coordinates $x_{1}$. Instead the  generalized eigenfunctions
corresponding to the other intervals decay  exponentially in $x_{1}$,
being of the Bloch-Floquet form in the longitudinal coordinate
$x_{2}$.
Such a type of surface states
(see Definition \ref{d31} below) were first found by Rayleigh in the problem
of oscillation of an homogeneous elastic half-space (see e.g.
\cite{Lo}), and since then were found and studied in a number of
problems, described by differential and finite difference
equations whose coefficients are strongly varying in coordinates $x_{1}$
(see e.g. \cite{KP} for a list of references on
respective physics results and applications). All these results
concerned the problems where coefficients were
constant in the  $x_{2}$ coordinates.
We analyze also the case where $\alpha _{n}=p_{n}/q_{n}$
approaches an irrational $\alpha $ as $n\rightarrow \infty $, and
we show that there exists a certain continuity of the spectrum in
this asymptotic regime. In particular, the width of surface bands,
lying outside of \ $[-d,d]$ is exponentially small in $q_{n}$ as
$n\rightarrow \infty $, and the bands approach the dense set of
eigenvalues, found in \cite{KP}.

%In this paper we give first a  transparent
%proof of the absolute
%continuity of the part $ \sigma(H) \cap [-d,d]$ of the spectrum
%$\sigma(H)$ of $H$ for arbitrary $d_1, d_2 \ge 1$,  and of the
%completeness of the wave operators of the pair $(H,H_0)$ for
%rationally independent $\alpha$'s. Then we find a complete system
%of generalized eigenfunctions $\Psi_{\pm}(\x,\kk), \x \in \ZZ^{d},
%\kk \in \TT^{d}$ having a ``Sommerfeld-like'' form. They are the sum of
%the incident plane wave of  unit amplitude  and of
%reflected or transmitted waves . We prove that $\Psi_{\pm}(\x,\kk)$ are the
%kernels of the wave operators $\Omega_{\pm}$. We report also
%certain results on the periodic two-dimensional case $d_1=d_2=1$,
%i.e. for $\alpha =p/q $.

%%%%%%%%%%%%%%%%%%%%%%%%%%%%%%%%%%%%%%%%%%%%%%%%%%%%%%%%%%%%%%%%%%%%%%%%%%%

\medskip
\section{Generalities}

\setcounter{equation}{0}
Recall that we are studying the self-adjoint operator $H$, acting
in $l^2(\ZZ^d)$ and defined in (\ref{ham}) - (\ref{par}).  The
operator is selfadjoint as the sum of the  multiplication
selfadjoint  operator $V$ of (\ref{Vt}), and  of the bounded
selfadjoint operator $H_0$ of (\ref{lap}). We will use an analogue
of the Cayley transform introduced in \cite{PF:84} for
 the "volume" potential $(d_1=0)$ and in \cite{KP} for the
"surface" case $(d_1=1)$, in both cases to study the pure point
spectrum for the Diophantine $\alpha's$ (see (\ref{di})).

\medskip \noindent To put the subsequent simple argument
in a more general context,
we rewrite the potential
 (\ref {Vt}) as
\begin{equation}\label{Vx}
V(\x)=v(x_2)\chi_S(\x),
\end{equation}
where $\chi_S$ is the indicator of the subspace $S=\ZZ^{d_2}$ and
we assume that $g>0 $ (the case  $g<0 $  can be treated
analogously). We define the orthogonal projection $P$ of $l^2(\ZZ^d)$ :
\begin{equation}\label{P}
(P\Phi)(\x)=\chi_S(\x)\Phi((0,x_2)),
\end{equation}
and we  write the potential (\ref{Vx}) in the form
\begin{equation}\label{Vs}
V= P v P,
\end{equation}
Here and in the following we use lower cases to denote operators acting 
on $l^2(S)$ defined by the restriction on $Pl^2(\ZZ^d)$ of 
the corresponding operator.

\medskip
\noindent We use as a starting point the well known
formulas for the resolvent $G(z)=(H-z)^{-1}$  of a selfadjoint
operator $H=H_0+V$:
\begin{equation}\label{GT}
G(z)=G_0(z)- G_0(z)T(z)G_0(z), \Im z \neq 0
\end{equation}
with
\begin{equation}\label{G0}
G_0(z)=(H_0-z)^{-1},  \quad T(z)=V-T(z)G_0(z)V.
\end{equation}
It follows from (\ref{Vs}) and from (\ref{G0}) that the operator $T(z)$ has the form:
\begin{equation}\label{TTP}
T(z)= P t(z) P,
\end{equation}
where the operator $t(z)$, acting on  $l^2(S)$, satisfies the
equation
\begin{equation}\label{t}
t(z)=v - t(z)\gamma_0(z) v,
\end{equation}
in which $\gamma_0(z)$ is defined from the restriction of $G_0(z)$ to the
subspace $Pl^2(\ZZ^d)$. The formal
solution of the equation is
\begin{equation}\label{Ts}
t(z)= v(1+  \gamma_0(z)  v)^{-1}= ( v^{-1}+  \gamma_0(z) )^{-1}.
\end{equation}
Let $u$ be the unitary operator in $l^2(S)$ defined by the relation:
\begin{equation}\label{u}
(u\psi)(x_2)=e^{-2i\pi \alpha.x_2}\psi(x_2), \quad x_2 \in S.
\end{equation}
Then, by using the Euler formula for the function $x \mapsto
\tan x$ and the notations above, we can write the potential
(\ref{vt}) as
\begin{equation}
v= \frac{g}{i} \cdot \frac{1-\sigma u}{1+\sigma u}, \label{v}
\end{equation}
where
\begin{equation}\label{sig}
\sigma= e^{-2i\pi \omega}.
\end{equation}

\medskip
\noindent Formulas (\ref{Vt}) - (\ref{sig}) motivate the following abstract
statement.
\begin{lemma}\label{l21}
Let $H$ be a selfadjoint operator, acting on
$l^2(\ZZ^d)$, and having the form $H=H_0+V$, where $H_0$ is a
selfadjoint operator  and $V$ is given by formulas
(\ref{Vs})and(\ref{v}) in which $S$ is any subset of $\ZZ^d$ and $|\sigma|\leq 1$.
 Define the following operators in $l^2(S)$
\begin{equation}\label{Ga0b}
b(z)=(g\gamma_0(z)-i)(g\gamma_0(z)+i)^{-1},
\end{equation}
assuming that $b(z)$ is bounded. If the operator $g\gamma_0(z)+i$
is invertible and if
\begin{equation}\label{b}
||b(z)||<1,
\end{equation}
then the operator $t(z)$, defined in (\ref{TTP}) and in (\ref{Ts}), can be
represented in the form:
 \begin{equation}\label{Tbu}
t(z)= g(1-\sigma u)(1-\sigma b(z) u)^{-1} (g\gamma_0(z)+i)^{-1},
\end{equation}
or in the form
\begin{equation}\label{qvers}
t(z)=g(g\gamma_0(z)+i)^{-1} \big[1-2i\sigma u
\sum_{l=0}^{q-1}(\sigma b(z)u)^l(1-(\sigma b(z)
u)^q)^{-1}(g\gamma_0(z)+i)^{-1}\big],
\end{equation}
where $\sigma$ is defined in (\ref{sig}), and $q \ge 1$ is an integer.
\end{lemma}

\noindent
\textit{Proof}. Note that the conditions $||b(z)||<1$ and
$|\sigma|\leq 1$ allow us to define the operator $(1-\sigma
b(z)u)^{-1}$  by the Neumann-Liouville series. Consider first the
case, where the modulus of the complex number $\sigma$ in
(\ref{v}) is strictly less than 1. In this case the operator
$(1+\sigma u)^{-1}$ is well defined and we obtain from
(\ref{v}), and from  (\ref{Ga0b}):
\begin{eqnarray*}
1+\gamma_0 v&=&\big[ i(1+\sigma u)+g\gamma_0(1-\sigma
u)\big]\big(i(1+\sigma u)\big)^{-1}\\ \nonumber &=&
(g\gamma_0+i)\left(1-(g\gamma_0-i)(g\gamma_0+i)^{-1}\sigma u
\right)\big(i(1+\sigma u)\big)^{-1},
\end{eqnarray*}
or $1+\gamma_0 v = (g\gamma_0+i) (1-b(z)\sigma u ) \big(i(1+\sigma
u)\big)^{-1}$, where the operators $\gamma_0(z)$, and $b(z)$ are
defined in (\ref{Ga0b}). Formulas (\ref{t}),
(\ref{v}), and the hypotheses of the lemma lead to (\ref{Tbu}) for
$|\sigma|<1$. According to inequality (\ref{b}) the
Neumann-Liouville series for $(1-b(z)\sigma u )^{-1}$ converges
for $|\sigma|=1$, and since the operator $(1+\sigma u)^{-1}$ is
not present in  formula (\ref{Tbu}), we can make the limit
$|\sigma| \to 1$ in the formula, proved for $|\sigma| <1$,
and obtain representation
(\ref{Tbu}) in the case $|\sigma|=1$.

\begin{proposition}\label{p21}
Let $H$ be the selfadjoint operator defined in Lemma \ref{l21} and
$G(z)=(H-z)^{-1}, \; \Im z >0$ be its resolvent. Assume that $z$
is such that the conditions of the Lemma \ref{l21} hold. Then
$G(z)$ can be represented as follows:
\begin{eqnarray}\label{Gs}
G(z)&=&G_0(z)- gG_0(z)P(g\gamma_0(z)+i)^{-1}PG_0(z)+
 2igG_0(z)P(g\gamma_0(z)+i)^{-1} \nonumber \\
& \times & \sigma u \sum_{l=0}^{q-1}(\sigma b(z)u)^l(1-(\sigma
b(z) u)^q)^{-1}(g\gamma_0(z)+i)^{-1}PG_0(z),
\end{eqnarray}
where $q \ge 1$ is an integer, $u$ is defined in (\ref{u})
and the operators
$\gamma_0(z)$, $b(z)$ are defined in
(\ref{Ga0b}).
\end{proposition}

\noindent {\it Proof.} The proposition follows easily from
(\ref{GT}), and from Lemma \ref{l21}.

\medskip
\noindent \textit{Remarks.} 1). In formula (\ref{v}) the unitary
operator $\sigma u$ can be viewed
as the Cayley transform of $v$ (see \cite{AG}
for the definition of the Cayley transform). Likewise, the
contraction operator $b(z)$ can be viewed as the Cayley transform
of the dissipative operator $i \gamma_0(z) \; (\Re i\gamma >0)$.
Hence, we can say that the passage from the operators $v^{-1}$ and
$\gamma_0(z)$ in (\ref{Ts}) to their Cayley transforms $\sigma u$
and $b(z)$ in the case of the potential (\ref{Vt}) - (\ref{vt})
leads to formulas (\ref{Tbu}) - (\ref{Gs}). This will allow us to
study the absolutely continuous spectrum of the operator $H$ for
any $d_1 \ge 0$, as it was done in papers \cite{PF} and \cite{KP}
for the pure point spectrum, despite that the subsequent
techniques to study the resolvent (\ref{Gs}) are different in
these two cases.

\smallskip
\noindent 2). Integrate formula (\ref{Gs}) with respect to $\omega
\in [0,1)$ and denote this operation by $\langle \cdots\rangle$.
We obtain: $$ \langle G(z) \rangle= G_0(z)- gG_0(z)P
(g\gamma_0(z)+i)^{-1}P G_0(z). $$ In view of the general formula
(\ref{Ts}), valid for any surface potential $v$, we can interpret
the equality $\langle t(z) \rangle= g(g\gamma_0(z)+i)^{-1}=
(-(ig)^{-1} + \gamma_0(z))^{-1}$ as the fact that $ \langle G(z)
\rangle$ is the resolvent of the Schr\" odinger operator  whose
surface potential is the complex constant $V(x)=-ig\chi_S(x)$.
This fact plays an important role in the interpretation of results
of analysis of the point spectrum of $H$ outside $[-d,d]$ in
\cite{KP}. Similar fact is known also in the case of the volume
potential (\ref{Vs}), i.e. for the case $S=\ZZ^d$ \cite{PF:84}.

\medskip \noindent Now we are going to show that the above
proposition is applicable
to the operator defined by (\ref{ham}) - (\ref{par}) where $S$ is chosen as $Z^{d_2}$ . 
To check the conditions of the lemma and the proposition
we will use the
Fourier transformation which we define as follows:
\begin{equation}\label{Ff}
\hat{\Phi}(\kk)= \sum_{\x \in \ZZ^{\nu}}e^{-2i\pi \x \cdot
\kk}\Phi(\x), \; \kk \in \TT ^\nu, \quad \Phi(x)=
\int_{\TT^{\nu}}d \kk e^{2i\pi \x \cdot \kk}\hat{\Phi}(\kk), \; \x
\in \ZZ^\nu,
\end{equation}
where $\TT^\nu=[0,1)^\nu$ is the $\nu$-dimensional unit torus.

\medskip \noindent By using  the Fourier transformation we
can write the following
representation of the Green function $G^{(\nu)}_0(\x-\y;z)$ of the
$\nu$-dimensional Laplacian (operator (\ref{lap}) for $d=\nu$):
\begin{equation}\label{Go}
G^{(\nu)}_0(\x-\y;z)= \int_{\TT^{\nu}} d \kk \frac{e^{2i\pi \kk
\cdot (\x-\y)}}{E_\nu(\kk)-z}, \; \Im z \neq 0,
\end{equation}
where
\begin{equation}\label{Ek}
E_\nu(\kk)=- \sum_{i=1}^{\nu} \cos 2 \pi k_i, \quad
(k_1,...,k_\nu)=\kk \in \TT^\nu.
\end{equation}

\medskip \noindent These formulas imply that the operator $\gamma_0(z)$ of
(\ref{Ga0b}) has the following matrix in $l^2(\ZZ^{d_2})$:
\begin{equation}\label{GaOG}
\gamma_0(x_2-y_2;z)=G_0^{(d)} ((0,x_2)-(0,y_2);z),
\end{equation}
i.e.  $\gamma_0(z)$ is a convolution operator in $l^2(\ZZ^{d_2})$.
In view of (\ref{Go}) we have:
\begin{equation}\label{GaO}
\gamma_0(x_2;z)= \int_{\TT^{d_2}}dk_2 e^{2i\pi k_2 \cdot
x_2}\int_{\TT^{d_1}} \frac{dk_1}{E_d(\kk)-z},
\end{equation}
or
\begin{equation}\label{GaxGak}
\gamma_0(x_2;z)= \int_{\TT^{d_2}}dk_2 e^{2i\pi k_2 \cdot x_2}G_0^{(d_1)}(0,z-E_{d_2}(k_2)).
\end{equation}
We  shall denote
\begin{equation}\label{Gag}
\hat \gamma_0(k_2;z): =G_0^{(d_1)}(0,z-E_{d_2}(k_2)),
\end{equation}
i.e. $\hat \gamma_0(k_2;z)$ is the symbol, representing the operator
$\gamma_0(z)$ in $L^2(\TT^{d_2})$ as a multiplication
operator. These formulas allow us to show that the hypotheses of
Lemma \ref{l21} and  Proposition \ref{p21} are valid for any $z,
\; \Im z >0$ (see Lemma \ref{lA2}). Besides, we have

%%%%%%%%%%%%%%%%%%%%%l22%%%%%%%%%%%%%%%%%%%%%%%%%%%%%%%%%%%%%%%%%

\begin{lemma}\label{l22}
Let $b(z)$ and $u$ be  the operators, defined by (\ref{Ga0b}) and
(\ref{u}). Then for any integer $m\geq 1$,
\begin{equation}\label{ub}
 \widehat {((b(z)u)^m \varphi)}(k_2)=\sigma^m
 \big(  \prod_{l=0}^{m-1} \hat b(k_2+l\alpha ;z)\big)
\hat \varphi(k_2+m\alpha), \quad k_2 \in \TT^{d_2},
\end{equation}
where $ \hat \varphi$ denotes
the Fourier transform of $\varphi\in l^2(\ZZ^{d_2})$ and
\begin{equation}\label{bGak}
\hat b(k_2;z)= \frac{g \hat \gamma_0(k_2;z)-i}{g \hat
\gamma_0(k_2;z)+i}
\end{equation}  where $\hat \gamma_0(k_2,z)$ is defined in
(\ref{Gag}).
\end{lemma}

%%%%%%%%%%%%%%%%%%%%%%%%%%%%%%%%%%FIN du l22%%%%%%%%%%%%%%%%%%%%%%%%%%%%%%%%%

\noindent {\it Proof.} It follows from (\ref{u}) that the operator
$u$ is the shift by $\alpha$ in the space $L^2(\TT^{d_2})$:
\begin{equation}\label{ushif}
\widehat{(u\varphi)}(k_2)=\hat \varphi(k_2+ \alpha).
\end{equation}
From this and the fact that $b(z)$ of (\ref{Ga0b}) is the
multiplication by the function $\hat b(k_2;z)$ of (\ref{bGak})
in the space $L^2(\TT^{d_2})$ prove the lemma.

\medskip \noindent We will obtain now a representation of the Green function of $H$
which will be central in the subsequent spectral analysis of the
absolutely continuous spectrum of the operator.

%%%%%%%%%%%%%%%%%%%%%%%THEOREM21%%%%%%%%%%%%%%%%%%%%%%%%%%%%%%%%%%%%%%%%%
\begin{theorem}\label{t21}
Let  $H$ be  the operator,  defined by (\ref{ham})-(\ref{par}. Then
the Green function of $H$ (the matrix in $l^2(\ZZ^d)$ of its
resolvent $G(z)=(H-z)^{-1}$) can be written in the form:
\begin{eqnarray}\label{GF}
G(\x, \y;z)&=&G_0^{(d)}(\x-\y;z) +\sum_{m=0}^\infty \int_{\TT^{d_2}} dk_2
e^{2i\pi k_2 \cdot (x_2-y_2)}t_m(k_2;z)
\nonumber
\\ &\times & G_0^{(d_1)}(x_1;z-E_{d_2}(k_2))
G_0^{(d_1)}(y_1;z-E_{d_2}(k_2+m\alpha))e^{-2i\pi m\alpha \cdot
y_2} ,
\end{eqnarray}
where
\begin {equation}\label{tm}
t_m(k_2;z)=\frac{g}{g \hat \gamma_0(k_2;z)+i} \left\{
\begin{array}{ll} \displaystyle -1, & m=0 \\ 2i\sigma (g \hat
\gamma_0(k_2 +\alpha;z)+i)^{-1}, & m=1 \\ 2i\sigma^{m+1} (g \hat
\gamma_0(k_2+m\alpha;z)+i)^{-1}\prod_{l=0}^{m-1} \hat
b(k_2+l\alpha ;z), & m\geq 2,
\end{array}
\right.
\end{equation}
$ G_0^{(d_1)}(x_1;z)$ is the Green function (\ref{Go}) of the
$d_1$-dimensional Laplacian, $E_{d_2}(k_2)$ is defined in
(\ref{Ek}) for $\nu=d_2$, and $\hat \gamma_0(k_2;z)$, $\hat
b(k_2;z)$ are defined respectively in (\ref{Gag})and (\ref{bGak}).

\medskip \noindent Besides, the (generalized) kernel  of the operator $T(z)$ of
(\ref{GT}) and of Lemma \ref{l21} has the following form in
$L^2(\TT^d)$:
\begin{equation}\label{Tkp}
T(\kk,\p;z)= \sum_{m=0}^\infty t_m(k_2;z) \delta(k_2+m\alpha-p_2),
\end{equation}
where $t_m(k_2;z)$ is defined in (\ref{tm}). In particular, the
kernel is independent of the components $k_1,p_1 \in \TT^{d_1}$ of
its arguments $\kk,\p \in \TT^{d}$.
\end{theorem}
%%%%%%%%%%%%%%%%%%%%%%%FIN du theorem21%%%%%%%%%%%%%%%%%%%

\medskip \noindent {\it Remark.} Formulas (\ref{GF}) and  (\ref{Tkp}) have
to be compared with the formulas for respective quantities for
point potential: $V(\x)=v\delta(\x), \ (d_2=0)$ \, and for the
constant surface potential: $V(\x)=v\delta(x_1), \
v=\mathrm{const}$. In the first case we have:
\begin{equation}\label{Gpo}
G(\x, \y;z)=G_0^{(d)}(\x-\y;z) - \frac{v}{1+vG_0^{(d)}(0;z)}
G_0^{(d)}(\x;z) G_0^{(d)}(\y;z),
\end{equation}
and
\begin{equation}\label{Tpo}
T(\kk,\p;z)=\frac{v}{1+vG_0^{(d)}(0;z)},
\end{equation}
while in the second case:
\begin{eqnarray}\label{Gco}
G(\x, \y;z)&=& G_0^{(d)}(\x-\y;z) - v\int_{\TT^{d_2}}dk_2
\frac{e^{2i\pi k_2 \cdot (x_2-y_2)}}
{1+vG_0^{(d_1)}(0;z-E_{d_2}(k_2))} \nonumber \\ & \times
&G_0^{(d_1)}(x_1;z-E_{d_2}(k_2))
 \  G_0^{(d_1)}(y_1;z-E_{d_2}(k_2)),
\end{eqnarray}
and
\begin{equation}\label{Tco}
T(\kk,\p;z)=\frac{v \delta(k_2-p_2)}
{1+vG_0^{(d_1)}(0;z-E_{d_2}(k_2))}.
\end{equation}
In particular the term, corresponding to $m=0$ in (\ref{GF}),
coincides with the second of (\ref{Gco}) in which $v$ is replaced
by $ig$.
%%%%%%%%%%%%%%%%%%%%%%%%%%%%%%%%%%%%%%%%%%%%%%%%%%%%%%%%%%%%%%%%%%%

\noindent {\it Proof of Theorem (\ref{t21})}.  According to
(\ref{b}), $\Vert  b(z)\Vert < 1$ if $\Im z \not= 0$. Hence we
can write the operator $(1-\sigma bu)^{-1}$ in  (\ref{Tbu}) for
$q=1$ as the Neumann-Liouville series in powers of $\sigma bu$.
Applying lemma \ref{l21} to each term of the series , we get
(\ref{GF}) after simple  algebra. Formula (\ref{Tkp}) follows from
(\ref{GT}) and (\ref{GF}). Theorem \ref{t21} is then proved.

\medskip  \noindent
\textit{Remark.} Formulas  (\ref{GF}) and  (\ref{Tkp}) are the
basic tools of spectral and scattering analysis of the operator
(\ref{ham}) presented in this paper. An advantage of these formulas
is that they are valid for all values of the spectral parameter
$z=E+i\varepsilon$, up to the real values $z=E \pm i0$, for $\vert E
\vert <d $, in the case of $\alpha $'s with rationally independent
components (quasi-periodic in $x_2$ potential $V(\x))$ and they
are valid for all
$E\in \mathbb{R}$ in the case of $\alpha$'s with rational components
(periodic in $x_2$ potential $V(\x))$.

\medskip \noindent One more general fact, concerning the
operator $H$ and necessary
in the sequel, is given by
%%%%%%%%%%%%%%%%%%%%%%%%%%thm22%%%%%%%%%%%%%%%%%%%%%%%%%%%%%%%%%%%%%%%%%%%%%
\begin{theorem}\label{t22}
Let $H=H_0+V$ be  the operator defined by (\ref{ham}), (\ref{lap})
and (\ref{Vx}). Then its spectrum $\sigma(H)$ contains the
interval $[-d,d]=\sigma(H_0)$ for all $g \in \R, \; \alpha \in
\R^{d_2}$ and $ \omega \in [0,1]$.
\end{theorem}
%%%%%%%%%%%%%%%%%%%%%%%%%FIN%%%%%%%%%%%%%%%%%%%%%%%%%%%%%%%%%%%%%%%%%%%%%%%%%
\noindent {\it Proof.} We will apply the H. Weyl criterion,
according to which $E \in \mathbb{R}$ belongs to the spectrum of a
self-adjoint operator $H$ if and only if there exists a sequence
$\{\Psi_n\}_{n\in \NN}$ of vectors of respective Hilbert space
such that  $\Vert \Psi_n\Vert=1 $, and that $\lim_{n\to
\infty}\Vert (H-E)\Psi_n\Vert=0 $.

\medskip
\noindent Denote by $ \mathbf{1}_r$ the indicator of the ball $\{
\x \in \ZZ^{d}: \; \vert \x\vert \leq r \}$ and set for all $\kk
\in \TT^{d}$,
 $$\Psi_n(\x)=\mathbf{1}_n(\x)(1- \delta(x_1))
e^{2i\pi \kk.\x}/ N_n ; \quad N_n^2= \sum_{\x\in \ZZ^d} \vert
\mathbf{1}_n(\x)(1- \delta(x_1)) \vert ^2 =
 O(n^d), \quad n \to \infty. $$
It is easy to  find that
\[
(H\Psi_n)(\x)= \left\{ \begin{array}{ll} E_d(\kk)\Psi_n(\x),&
\vert \x\vert \leq n-2, \vert x_1\vert \geq 2;\\ A_n(\x), & n-2
\leq \vert \x\vert \leq n+2;\\ b_n(\x), & \vert x_1\vert \leq 1;\\
0, & \vert \x\vert \geq n+3,
\end{array}
\right.
\]
where $\Vert A_n \Vert = O(n^{-1/2})$, $ \Vert b_n\Vert =
O(n^{-d_1/2})$ as $n\to \infty$. This proves the theorem.

%%%%%%%%%%%%%%%%%%%%%%%%%%%%%%%%%%%%%%%%%%%%%%%%%%%%%%%%%%%%%%%%%%%%%%%%%%%%%%%%%%%%%%%%%%%%%%%%%%
%%%%%%%%%%%%%%%%%%%%%%%%%%%%%%%%%SECTION3%%%%%%%%%%%%%%%%%%%%%%%%%%%%%%%%%%%%%%%%%%%%%%%%%%%%%

\medskip
\section{Absolute Continuous Spectrum in the Almost Periodic Case}

\setcounter{equation}{0}
In this section we assume that the vector $\alpha \in \R^{d_2}$
from (\ref{Vt})  has rationally independent components, i.e. that
the relation $\alpha_1 r_1+ ...+\alpha_{d_2} r_{d_2}=0$ with
rational coefficients $r_1,...,r_{d_2}$ implies that all these
coefficients are equal to zero.

%%%%%%%%%%%%%%%%THEOREM31%%%%%%%%%%%%%%%%%%%%%%%%%%%%%%%%%%%%%%%%%%%%%%%%%%%

\begin{theorem}\label{t31}
 Let  $H=H_0+V$ be the self-adjoint operator defined by
(\ref{ham}) - (\ref{par}) in which  the vector $\alpha \in
\R^{d_2}$ has rationally independent components. Then  $H$
has purely absolutely continuous spectrum on the interval $(-d,d)$.
\end{theorem}
%%%%%%%%%%%%%%%%%%%%%%%%%%%%%%%%%%%%%%%%%%%%%%%%%%%%%%%%%%%%%%%%%%%%%%%%%%%%%%

\noindent {\it Proof.} According to the general principles (see e.g.
\cite{RS}), it suffices to prove that for any vector $ \Phi \in
l^2(\ZZ^{d})$ of a dense set the limit $\Im (G(E+i0)\Phi,\Phi)$
exists and is bounded for all $E \in (-d,d)$. Restricting
ourselves to the vectors concentrated at a point $ \x \in
\ZZ^{d}$, i.e.  to the vectors $\delta_{\x}=\{\delta (\x-\y)\}_{\y
\in \ZZ^{d}}$, we have to prove that  for any $\x \in \ZZ^{d}$ the limit
$\Im G(\x,\x;E+i0)$ exists and is bounded for all $E \in (-d,d)$.
We shall prove more, namely that $G(\x,\y;E+i0)$ exists and is bounded
for all $E \in (-d,d)$ and all $\x,\y \in  \ZZ^{d}$. In view of Theorem
\ref{t21}, we have to prove that the series of (\ref{GF})
converges  not only for $\Im z>0$ but also for $\Im z=0$.

\medskip \noindent Since the vector $\alpha$ has rationally independent
components, we have uniformly in $ k_2 \in \TT^{d_2}$ and for any
$\gamma >0$ (see e.g. \cite{FKS}):
\begin{equation}\label{erg0}
\lim_{m \to \infty} \sharp \{l \in \ZZ:  \; k_2 + l \alpha \in
K_\gamma (E), \;  1\leq l \leq m \} m^{-1}=
 \vert K_\gamma (E)\vert,
\end{equation}
where
\begin{equation}\label{Kga}
K_\gamma (E)= \{k_2 \in \TT^{d_2}: \; E- E_{d_2}( k_2) \in [-d_1
+\gamma,d_1 -\gamma ]\},
\end{equation}
and $ \vert K_\gamma (E)\vert$ denotes the Lebesgue measure of the set
 $K_\gamma(E) \subset \TT^{d_2}$. It is easy to check that for any
$\vert E\vert<d$ there exists $\gamma>0$ such that $K_\gamma(E)$
is an open set of $ \TT^{d_2}$. According to Lemma \ref{lA3}, in
this case there exists $\delta>0$ such that $  \vert  \hat b (k_2,
E+i0)\vert  \leq 1-\delta, \forall k_2 \in K_\gamma (E)$, and
according to (\ref{erg0}), there exists $ m_0> 0$ such that
\[
\sharp \{l \in \ZZ:  \; k_2 + l \alpha \in K_\gamma (E), \; 1\leq
l \leq m \} \geq \frac{ m}{2} \vert K_\gamma (E)\vert
\]
for all $ m\geq m_0$. Hence we have the following bound
for the product in the r.h.s. of (\ref{GF}):
\begin{equation}\label{bou}
\left|\prod_{l =0}^{m-1} \hat b (k_2+l\alpha; E+i0)\right| \leq
(1-\delta)^{ m\vert K_\gamma (E)\vert/2}, \quad m \geq m_0,
\end{equation}
and the series in the r.h.s. of (\ref{GF}) converges uniformly in
$k_2 \in \TT^{d_2}$. Besides, by using bound (\ref{bou}) and Lemma
\ref{lA5}, it can be shown that for
 $\vert E \vert \leq d-\gamma, \; \gamma > 0 $, the
series is bounded  in $k_2 $ and $E$ , hence we can
integrate the series with respect to $k_2$. Theorem is proved.
%%%%%%%%%%%%%%%%%%%%%%%%%%%%%%%% REMARKS%%%%%%%%%%%%%%%%%%%%%%%%%%%

\medskip \noindent {\it Remarks.} 1). Another form to express (\ref{erg0}) -
(\ref{bou}) is to write the relation:
\begin{equation}
\lim_{m\to \infty} \left| \prod_{l =0}^{m-1} \hat b (k_2+l\alpha;
E+i0) \right|^{1/m} = \exp\left\{ \int_{\TT^{d_2}}dq_2 \log \vert\hat b
(q_2; E+i0)\vert \right\}, \label{erg1}
\end{equation}
valid uniformly in $k_2 \in \mathbb{T}^{d_2}$ (see \cite{FKS}) and
showing that if $\vert E\vert \leq d-\gamma,  \gamma >0 $, then
the integral in the r.h.s. is negative, thus the product in the
l.h.s. is exponentially decaying in $m$ as $m \to \infty$.

\smallskip \noindent 2). Theorem \ref{t31} reveals a fairly simple mathematical
mechanism responsible for the absolutely continuous spectrum for
the "subspace" potential (\ref{Vt}) -  (\ref{vt}) with $d_1\geq 1$
(recall that in the "volume" case $d_1=0,  d_2=d$, the absolutely
continuous spectrum is absent, moreover if $ \alpha$ is
Diophantine then the spectrum is pure point \cite{PF:84}). The
mechanism is the positiveness of the imaginary part of $ \hat
\gamma_0(k_2;E+i0)= G_0^{(d_1)} (0,E+i0-E(k_2) )$ in a certain
domain of $(E,k_2)$. This is most transparent in the "genuine
surface" case $d_1=1$, where $ G_0^{(1)}(0,E+i0 )$ is pure
imaginary if $ \vert E \vert <1$ and is pure real if $\vert E
\vert \geq 1$,  (see formula (\ref{G01}) below). In the latter
case $ \vert\hat b (k_2; E+i0)) \vert =1 $ and the series
(\ref{GF}) diverges for a dense set of energies (see \cite{KP}).
This leads to the pure point spectrum everywhere outside
of the spectrum $ \sigma(H_0) $ of the Laplacian (similarly to the volume
case \cite{PF:84},
 where the analogue of $\hat \gamma_0(k_2;E)$ in  (\ref{bGak})
is real for all $ E \in \R $). In the former case $ \vert\hat b
(k_2; E+i0)) \vert  $ is strictly less than $1$ for any $ E \in
(-d,d)$ on an open set of $k_2 \in \TT^{d_2}$, the series is
convergent and the spectrum inside of $\sigma(H_0)=[-d,d]$ is pure
absolutely continuous.

\medskip \noindent
As usual in scattering theory, a fact of primary interest is the
existence and completeness of wave operators
 $ \Omega_{\pm}=\mathrm{s}\cdot\lim_{t\to \pm \infty}e^{itH}e ^{-itH_0}
\mathcal{E}_0( \Delta)$, where $ \mathcal{E}_0$ is the resolution
of identity of $H_0$, and $ \Delta$ is an interval of the spectral
axis. In the next theorem we prove these properties in our case.

\medskip \noindent We mention first that in papers
\cite{JL1,JL2,JM1,JM4}, the
scattering theory was developed for the operator $ H_1$, acting in
 $l^2(\ZZ_+^d)$, $\ZZ_+^d=\{ (x_1,x_2) \in \ZZ^d; x_1 \geq 0,
x_2 \in \ZZ^{d-1}\}$, and defined as:
\begin{equation}\label{H1}
(H_1\Psi)(\x)= \left\{ \begin{array}{l} \sum_{ \vert \x-\y
\vert=1} \Psi(\y), \; x_1 \geq 1;\\ \Psi(1,x_2)+  \sum_{ \vert
x_2-y_2 \vert=1} \Psi(0,y_2) + v(x_2)\Psi(0,x_2), \quad x_1 = 0
\end{array}
\right.
\end{equation}
for certain random and almost periodic surface potentials $v$. The
operator can be viewed as a boundary value problem
for the discrete Laplacian in $l^2(\ZZ_+^d)$ with the boundary
condition $\Psi(-1,x_2)= v(x_2)\Psi(0,x_2),  x_2 \in \ZZ^{d-1}$.
The ``unperturbed'' operator $H_0$ here is the discrete Dirichlet
Laplacian, corresponding to $v \equiv 0$ in (\ref{H1}). The
operator $H_1$ is closely related to our operator $H$ of
(\ref{ham}) for the surface case $d_1=1, d_2=d-1$ via standard
Green's formulas.
%%%%%%%%%%%%%%%%%%%%%%%%%%%%%thm32%%%%%%%%%%%%%%%%%
\begin{theorem}\label{t32}
Under the conditions of the Theorem \ref{t31}, the  wave operators
$\Omega_\pm$ for the pair $(H,H_0)$, defined by (\ref{ham})-
(\ref{par}), exist and are complete for any closed interval $
\Delta= [a,b] \subset (-d,d)$.
\end{theorem}
%%%%%%%%%%%%%%%%%%%%%%%%%%%%%%%%finthm%%%%%%%%%%%%%%%%%%%%%%%%

\noindent {\it Proof.} Existence of wave operators
 is a rather general fact. It was proved in \cite{JL1} for
a general surface perturbation $v $ in (\ref{H1}). In our case the
proof is practically the same. Thus we have to prove the
completeness. Mimicking the argument of \cite{JM1,JM4}, developed
for the boundary value problem (\ref{H1}), it is easy  to reduce
the proof of completeness to the proof of the relation:
\begin{equation}\label {sup1}
\sup_{\varepsilon >0 ,E\in [a,b]} \sum_{x_2 \in \ZZ^{d_2}} \vert
G((x_1,x_2) ,\y; E \pm i\varepsilon) \vert^2 < \infty
\end{equation}
for any  fixed $x_1 \in \ZZ^{d_1}, \y \in \ZZ^{d}$ and $ [a,b]
\subset (-d,d)$. Our formulas (\ref{GF}) - (\ref{tm}) for the
Green function of $H$ can be written in the form:
\[
G((x_1,x_2),\y;z)= \int_{\TT^{d_2}}dk_2 e^{2i\pi k_2 \cdot
x_2}G((x_1,k_2),\y;z),
\]
where
\begin{eqnarray}\label {Gkx1}
G((x_1,k_2),\y;z)& = & G_0^{(d_1)}(x_1-y_1;z-  E_{d_2}(k_2))-
\sum_{m=0}^{\infty}t_m(k_2,z) G_0^{(d_1)}(x_1;z-E_{d_2}(k_2))
\nonumber \\ &\times &
G_0^{(d_1)}(y_1;z-E_{d_2}(k_2+m\alpha))e^{2i\pi y_2 \cdot ( k_2+
m\alpha)}.
\end{eqnarray}
Thus, applying the Parseval equality for the Fourier transform
with respect to the variable $x_2$, we can present the sum in
(\ref{sup1}) as:
\begin{equation}\label {sup2}
\int_{\TT^{d_2}}dk_2  \vert G((x_1,k_2),\y;E+i\varepsilon)\vert^2.
\end{equation}
We have shown in the proof of Theorem \ref{t31} that the series
(\ref{Gkx1}) converges uniformly in $k_2 \in \TT^{d_2}$ for
$z=E+i\varepsilon, E \in [a,b] \subset (-d,d); \varepsilon> 0$.
Hence the integral in (\ref{sup2})  is finite for these values of
$ E$ and $\varepsilon$. This proves (\ref{sup1}).

\medskip \noindent In the next theorem we construct a family of generalized
eigenfunctions of $H$, relating them to the Green function of the
operator, as in the conventional scattering theory \cite{Pe,Si}.
%%%%%%%%%%%%%%%%%%%%%%%%THEOREM33%%%%%%%%%%%%%%%%%%%%%%%%%%%%%%%%%%%%%%%%%

\begin{theorem}\label{t33}
 Let  $G(\x,\y;z)$ be the Green function of the  operator
 $H=H_0+V$,
 defined by (\ref{ham}) - (\ref{par}), in which the vector
 $\alpha$ is rationally independent. Set
\begin{equation}\label {Gk}
G(\x,\kk;z)= \sum_{\y \in z^d}G(\x,\y ;z)e^{2i\pi\kk \cdot\y},
\quad \kk \in \TT^d,
\end{equation}
\begin{equation}\label {pSo}
\Psi_{z}(\x,\kk)= (E_d(\kk) -z)G(\x,\kk;z),
\end{equation}
and \begin{equation} \label{Tdot}
\dot \TT^{d_2}= \TT^{d_2}
\setminus \{{\stackrel {\mathrm{d_2-times}}
 {\overbrace {(0,0,...,0)} }},
{\stackrel  {\mathrm{d_2-times}} { \overbrace {(\pi,\pi,...,\pi)}
}}\}; \quad \dot \TT^{d}=
 \TT^{d_1}\times \dot \TT^{d_2}.
 \end{equation}
Then,   for  $z=E_d(\kk)\mp i \varepsilon $, the  limits:
\begin{equation}\label {lim}
\Psi_{\pm}(\x,\kk) = \lim_{\varepsilon \to +0} \Psi_{z}(\x,\kk)
\mid_{z=E_d(\kk) \mp i\varepsilon} =
 \lim_{\varepsilon \to +0} \pm i\varepsilon G(\x,\kk;E_d(\kk) \mp i\varepsilon),
\end{equation}
exist for all  $ \kk \in \dot \TT^{d}$, are bounded in  $ \x \in
\ZZ^d$ for any
  $ \kk \in \dot \TT^{d}$, are continuous in $\kk$ varying
in any compact set of $ \dot \TT^{d}$, and have the form
\begin{equation}\label{Som}
 \Psi_\pm(\x,\kk)= e^{2i\pi\kk \cdot \x}
 +\sum_{m=0}^\infty t_m(k_2-m\alpha;z)
G_0^{(d_1)}(x_1;z- E_{d_2}(k_2-m\alpha ))\Big|_{z=E_d(\kk) \mp i0}
e^{2i \pi (k_2-m\alpha) \cdot x_2},
\end{equation}
where the coefficients $t_m(k_2,z)$ are defined in (\ref{tm}).

\medskip \noindent Moreover:

\medskip \noindent {\it (i)}  the functions $\Psi_{\pm}(\x,\kk)$
satisfy the Schr\"odinger equation
in $\x$ for any $\kk \in \dot \TT^{d}$:
\begin{equation}\label{SEK}
((H_0 + V) \Psi_{\pm})(\x,\kk)= E_d(\kk) \Psi_{\pm}(\x,\kk);
\end{equation}

\smallskip \medskip \noindent  {\it (ii)}  the functions $\Psi_{\pm}(\x,\kk)$
are the unique solutions of the equation:
\begin{equation}\label{LS}
\Psi_{\pm}(\x,\kk)= e^{2i\pi\kk\cdot \x} - \sum_{\y \in
\ZZ^d}G_0^{(d)}(\x-\y; E_d(\kk) \mp i0)V(\y)\Psi_{\pm}(\y,\kk).
\end{equation}
 for any $\kk \in \dot \TT^{d}$
in the class of sequences $\Psi= \{\Psi(\x)\}_{ \x \in \ZZ^d}$
whose restrictions $\psi=\{ \Psi(0,x_2)\}_{ x_2 \in \ZZ^{d_2}}$
and the sequences $ \{(1+ \sigma e^{-2i\pi\alpha \cdot x_2})
\psi(x_2)\}_{ x_2 \in \ZZ^{d_2}}$ are representable as the Fourier
transforms of measures of bounded variation on $\TT^{d_2}$, and
the  sum of the r.h.s. of (\ref{LS}) is understood as the
generalized convolution of respective functions and measures;

\smallskip \noindent {\it (iii)} the families $ \{\Psi_\pm (\cdot,\kk)\}_{ \kk
\in \dot{\TT}^d}$ are orthonormalized, i.e. if for any continuous
function $ \hat{\Phi}$ of compact support in $\dot{\TT}^d$ we
set:
\begin{equation}\label{psifi}
 \Phi_\pm(\x)=\int_{\dot{\TT}^d}\Psi_\pm(\x,\kk) \hat{\Phi} (\kk)d\kk,
\end{equation}
then for any two such functions $\hat{\Phi}^{(1)}$ and
$\hat{\Phi}^{(2)}$ we have:
\begin{equation}\label{ort}
 \sum_{\x \in \ZZ^d} \Phi^{(1)}_\pm(\x) \overline{\Phi^{(2)}_\pm(\x)}
=\int_{\dot \TT^d}d\kk \hat{\Phi}^{(1)}(\kk)
\overline{\hat{\Phi}^{(2)}(\kk)};
\end{equation}
\medskip \noindent {\it iv)} the functions $\Psi_\pm :\ZZ^d \times
\dot{\TT}^d \to \CC$ are the kernels of the wave operators
$\Omega_\pm$, whose existence and completeness are proved in
Theorem \ref{t32}, i.e. for any $ \Phi \in l^2(\ZZ^d)$ such that
the support of its Fourier transform $ \hat{\Phi}$ is a compact
set in $\dot \TT^d$ we have:
\begin{equation}\label{Ompsi}
 (\Omega_\pm \Phi)(\x)=\int_{\dot{\TT}^d} \Psi_\pm(\x,\kk)
 {\hat \Phi}(\kk)d\kk.
\end{equation}
\end{theorem}
%%%%%%%%%%%%%%%%%%%%%%%%%%%%finthm%%%%%%%%%%%%%%%%%%%%%%%%%%%%%%%%%%%%%%%

\noindent {\it Proof.} We use again our basic formulas (\ref{GF})
- (\ref{tm}) for the resolvent of $H$. Making the Fourier
transform of (\ref{GF}) with respect to $\y$ and multiplying the
result by $ E_d(\kk)-z$, we present  (\ref{pSo}) in the form:
\begin{equation}\label{psiz}
 \Psi_z(\x,\kk)= e^{2i\pi\kk \cdot \x}
 + \sum_{m=0}^\infty t_m(k_2-m\alpha;z)
G_0^{(d_1)}(x_1;z- E_{d_2}(k_2-m\alpha))e^{2i\pi
(k_2-m\alpha)\cdot x_2}.
\end{equation}
Each term in this series  is continuous in $k_2$ and
$z,\Im z
>0$ and can be extended to real $z=E +i0, E \in [a,b]$, if the
closed interval $[a,b]$ lies strictly inside $(-d,d)$. According
to bound (\ref{bou}), the series converges uniformly in $k_2 \in
\dot \TT^{d_2}$ and $a\leq \Re z\leq b, \Im z \geq 0$, hence it
defines a continuous  function in this domain. This allows us to
perform the limits  (\ref{lim}) for $k_2 \in \dot \TT^{d_2}$ and
to obtain formula (\ref{Som}).

\medskip \noindent Our limitation
$\kk \in \dot \TT^{d}$, where $\dot \TT^{d}$ is
defined in (\ref{Tdot}) is necessary because for $\kk \in \TT^{d}
\setminus \dot \TT^{d}$ and for the respective two values of
$E=\pm d$ we cannot guarantee the validity of bound (\ref{bou}),
thus the convergence of the series in formula (\ref{Som}).

\medskip \noindent Let us prove now property (i) of $\Psi_\pm(\x,\kk)$.
We have obviously:
$$\sum_{ \vert \T-\x \vert =1} H_0(\x-\T)
G(\T,\y;z)-EG(\x,\y;z) + V(\x)G(\x,\y;z)=i\varepsilon G(\x,\y;z) +
\delta(\x-\y).
$$
The definition  (\ref{pSo}  of $ \Psi_z(\x,\kk)$
 and an easy justification of the interchange of the
multiplication by $V(\x)$ and of the Fourier transformation in
$\y$ in the third term of l.h.s. lead to  the equality:
\[\sum_{\vert \T-\x \vert =1}H_0(\x-\T) \Psi_z(\T,\kk)-E\Psi_z(\x,\kk) +
V(\x)\Psi_z(\x,\kk)=i\varepsilon (\Psi_z(\x,\kk)+ e^{2i\pi\kk
\cdot\x} ).
\]
Now, in view of relation (\ref{lim}), the limit of the r.h.s. of
the last equality is zero as $ \varepsilon \to 0$, and we get
(\ref{SEK}).

\medskip \noindent Let us prove now assertion (ii)
of the theorem, i.e. that
$\Psi_\pm(\x,\kk)$ satisfy (the Lippmann-Schwinger) equation
(\ref{LS}). We remark first that any solution $\Psi$ of (\ref{LS})
is uniquely determined by its restriction
$\psi(x_2)=\Psi((0,x_2))$ to the subspace  $\ZZ^{d_2}$, and that
$\psi$ verifies the equation, that can be symbolically written as:
\begin{equation}\label{LS2}
 \psi (x_2)= e^{2i\pi k_2 \cdot x_2} - (\gamma_0 v \psi)(x_2), \;
x_2 \in \ZZ^{d_2}.
\end{equation}
Hence we have to verify that the restriction $\psi(x_2,\kk)$ of
(\ref{Som}) to $\ZZ^{d_2}$ satisfies (\ref{LS2}). By using
(\ref{psiz} and (\ref{Som}), we can write the restriction
symbolically in the form:
\begin{equation}\label{psi21}
 \psi= \big\{1 - (g \gamma_0 +i)^{-1} g \gamma_0
[1-2i \sigma u(1-\sigma bu)^{-1} (g \gamma_0+i)^{-1} ] \big\}e_2
\Big|_{z=E_d(\kk)\mp i0},
\end{equation}
where $e_2(x_2)=e^{2i\pi k_2 \cdot x_2}$ and we used definition
(\ref{Ga0b}) of $\gamma_0$. The symbols $\gamma_0$, $b$ and $u$ in
the formula denote now not operators on $l^2( \ZZ^{d_2})$ or in
$L^2( \TT^{d_2})$, defined in (\ref{Ga0b})  and in (\ref{u}), but
just operations acting on sequences (functions of $ x_2 \in
\ZZ^{d_2} $) and representable as Fourier transforms of measures
of bounded variation depending on the parameter
$z=E(\kk)\mp i0, \; \kk \in \dot
\TT^{d_2}$. In order words, they belong to the
linear manifold:
\begin{equation}
{\cal L}_{\kk}= \{ f(x_2), x_2 \in \ZZ^{d_2}: f(x_2)=
\int_{\TT^{d_2}}  e^{2i\pi p_2 \cdot x_2} M_{\kk}(dp_2); \quad
\mathrm{Var}M_{\kk} < \infty\}. \label{space}
\end{equation}

\medskip \noindent The operations $b$ and $\gamma_0$ are multiplications of $M_{\kk}$
by ${\hat b} (p_2,z)$ and by ${\hat \gamma}_0(p_2,z)$ with
$z=E_d(\kk) \mp i0$, and $u$ is the shift by $ \alpha$ of the
measure. The operation $(1- bu)^{-1}$ is defined by the series $
\sum_{m=0}^\infty(bu)^m$, whose terms are given by (\ref{ub}), and
which converges for all $\kk \in \dot \TT^{d}$.
 By using these facts and a simple algebra, we can
rewrite (\ref{psi21}) as:
\begin{equation}
 \psi= i(1+\sigma u) (1-\sigma bu)^{-1} (g \gamma_0+i)^{-1}e_2
 \Big|_{z=E_d(\kk) \mp i0}. \label{psi22}
\end{equation}
Hence we have for the r.h.s. of (\ref{LS2}:
$$ e_2-\gamma_0 v
\psi=  e_2 - g \gamma_0 (1-\sigma u) (1+\sigma u)^{-1}(1+\sigma
u)(1-\sigma bu)^{-1} (g \gamma_0+i)^{-1}e_2
$$
or
$$ e_2 -\gamma_0 v \psi = \{1 - g \gamma_0 (1-\sigma u) (1-\sigma
bu)^{-1}(g \gamma_0+i)^{-1}\}e_2,
$$
meaning that the complex spectral parameter $z$ is replaced
by $E(\kk) \mp i0$. The r.h.s.
of the relation coincides with $\psi$. To prove this fact we have
to repeat the arguments leading to (\ref{Som}) and (\ref{psi21}),
but starting from formula (\ref{Tbu})  for the operator $T(z)$
instead formula (\ref{qvers}) . Thus we have proved  that (\ref{Som}) solves
(\ref{LS}).

\medskip \noindent Let us prove that (\ref{Som}) is the unique solution of
(\ref{LS}) in ${\cal L}_{\kk}$ and such that
their multiplication by $(1+\sigma e^{2\pi ik_2 \cdot x_2 })$
belongs also to ${\cal L}_{\kk}$.
Consider the homogeneous equation, corresponding to (\ref{LS}):
\begin{equation}\label{hLS}
\chi=\gamma_0 v \chi
\end{equation}
on the same manifold, and write the equality $\chi=(1+u) \varphi$,
where $\varphi$ also belongs to  (\ref{space}). Then we obtain the
following equation for $\varphi$:
\[ (1+g\gamma_0)(1-\sigma b u)
\varphi=0,
\]
where the symbols $\gamma_0$, $b$, and $u$ are again understood as
operations in the class ${\cal L}_{\kk}$. Applying to this relation
the operation $(1-\sigma bu)^{-1}(g\gamma_0+i)^{-1}$,which is well defined
in ${\cal L}_{\kk}$, we obtain: $\varphi=0$.

\medskip \noindent
According to the above considerations the second term in the r.h.s. of (\ref{LS})
 is the Fourier transform of the product of
$G_0^{(d_1)}(x_1;E_d(\kk)-E_{d_2}(k_2)+i0)$ (the Fourier
transform of $G_0^{(d)}(\x;z)\big|_{z=E_d(\kk+i0}$ in $x_2$) and of the
measure $M_{\kk}$, corresponding to $v\psi$:
\begin{equation}
\int_{\TT^{d_2}}G_0^{(d_1)}(x_1;E_d(\kk)-E_{d_2}(p_2)+i0)
M_{\kk}(dp_2). \label{rhs}
\end{equation}
In view of (\ref{v}), and  (\ref{psi22}) we have:
\begin{eqnarray}
v\psi&=&g(1-\sigma u)(1-\sigma b u)^{-1}(g\gamma_0+i)^{-1}e_2
\Big|_{z=E_d(\kk) \mp i0} \nonumber \\ &= &
g(g\gamma_0+i)^{-1}(1-2i\sum_{m=0}^{\infty}\sigma u (\sigma
bu)^{m} (g\gamma_0+i)^{-1}e_2 )\Big|_{z=E_d(\kk) \mp i0}.
\end{eqnarray}
By using this relation and the notations
introduced in Lemma \ref{l21} and in
Theorem \ref{t21}, we obtain that the measure  corresponding
to the second term of the r.h.s. of (\ref{LS}) is
\[
\sum_{m=0}^\infty t_m(k_2-m\alpha;E_d(\kk) \pm i0)
\delta(k_2-m\alpha-p_2).
\]
Combining these formulas we obtain
(\ref{LS}).

\medskip \noindent Let us prove now the orthogonality of $\Psi_\pm(\x,\kk)$,
 corresponding to different $\kk$'s,  i.e.
relation (\ref{ort}). It is clear that it is sufficient to prove
(\ref{ort}) for $ \Phi^{(1)}= \Phi^{(2)}$. The proof is rather
technical and we outline only its scheme, considering, say
$\Phi_-$.

\medskip \noindent The first step is the proof of the relation:
\begin{equation}\label{wave}
\lim_{\varepsilon \to + 0} \sum_{\x \in \ZZ^d} \vert  \Phi_-(\x) -
\Phi_\varepsilon(\x) \vert^2=0,
\end{equation}
where (cf (\ref{psifi}))
\begin{equation}\label{Phie}
\Phi_{\varepsilon}(\x)=\int_{\mathbb{T}^d}\Psi_{E_d(\kk)+i\varepsilon}(\x,\kk)
\hat{\Phi}(\kk)d\kk ,
\end{equation}
and  $\Psi_z(\x,\kk)$ is defined in (\ref{pSo}), i.e.
$\Psi_{E_d(\kk)+i\epsilon}(\x,\kk)= -i\varepsilon G(\x,\kk;
E_d(\kk) + i\varepsilon)$. The proof is based on formulas
(\ref{psiz}), and (\ref{Som}), and on the continuity of $
G_0^{(d)}(\x, E + i \varepsilon)$ with respect to $ \varepsilon >
0$. It is given in Lemma \ref{l31} below.

\medskip \noindent The second step is the proof of the relation:
\begin{equation}\label{L2no}
 \lim_{\varepsilon
\to + 0} \sum_{\x \in \ZZ^d} \vert \hat{\Phi}_\varepsilon(\x)
\vert^2 =\int_{\mathbb{T}^d} \vert \hat{\Phi}(\kk) \vert^2 d\kk,
\end{equation}
 which implies (\ref{ort}). We will just
sketch a proof of this relation.

\medskip \noindent Write the resolvent identity for the pair $ G(\bar z')$
and $ G(z'')$:
\begin{equation}\label{ri}
 \sum_{\T \in \ZZ^d} G(\T,\y; z'') \overline{ G(\T,\x;z')}
 =({\bar z'}-z'')^{-1}\big(G(\T,\x; {\bar z'}) - G(\T,\y; z'')\big).
\end{equation}
Replace in the r.h.s. of the identity $G $ by
$G_0 - G_0TG_0$ (see (\ref{GT})). We obtain after a simple algebra:
\begin{equation}\label{rid}
 G_0'G_0'' + ({\bar z'}-z'')^{-1}\big(G_0'T'G_0'-
 G_0''T''G_0''\big),
\end{equation}
where $ G_0'= (H_0 - {\bar z'})^{-1}$, $ G_0''= (H_0 - z'')^{-1}$
and $T'$ and $T''$ are the $T$-operators for the spectral
parameters ${\bar z'}$ and $z''$ respectively. Now we  make the
Fourier transformation with respect to $\x$ and $\y$, multiplying
(\ref{ri}) and (\ref{rid}) by $ e^{2i\pi\p \cdot \y -2i\pi\kk
\cdot \x}$ and summing the result over $\x, \y \in \ZZ^d$. The
l.h.s. of the obtained relation is $(\Psi_{z''}, \Psi_{z'})$. As
for the r.h.s., it can be written symbolically as:
\begin{equation}\label{dTE}
\delta(\kk-\p)-  T(\kk,\p;{\bar z'})\left(\frac{1}{ E_d(\p)-{\bar
z'}} - \frac{1}{z''-{\bar z'}}\right) -
 T(\kk,\p; z'')\left( \frac{1}{ E_d(\kk)-z''} -
 \frac{1}{{\bar z'}- z''}\right),
\end{equation}
where $T(\kk,\p;z)$ is the kernel in $L^2(\mathbb{T}^d)$ of the
$T$-operator, whose expression is given in Theorem \ref{t21}.
Setting in  (\ref{dTE}), $z'=E_d(\p)+ i \varepsilon $, and
$z''=E_d(\kk)+ i \varepsilon $, we obtain:
\begin{eqnarray*}
\delta(\kk-\p)&-&
\Big(T(\kk,\p;E_d(\kk)+ i \varepsilon ) +
 T(\kk,\p;E_d(\p)+ i \varepsilon)\Big)\\&\times&
\left( \frac{1}{ E_d(\p)-E_d(\kk)+ i \varepsilon} -
\frac{1}{E_d(\p)- E_d(\kk)+ 2i \varepsilon}\right).
\end{eqnarray*}
After multiplication by $ \hat{\Phi}(\kk) \overline{
\hat{\Phi}(\p)}$,
where $  \hat{\Phi}(\kk) $ is a smooth
function whose compact support is strictly inside $ {\dot
\TT}^{d}$, and after the subsequent integration with respect to $
\kk,\p \in \TT^{d}$, the second term of the last expression  tends
(weakly) to zero as $ \varepsilon \to 0$. We use the explicit form
(\ref{Tkp})  of the kernel $T(\kk,\p;z)$ to prove that $T(\kk,\p;E
+ i \varepsilon)$ is weakly bounded  in  $ \varepsilon \geq 0$, if
$\kk, \p$ are strictly inside of ${\dot \TT}^{d}$ and $ \vert E
\vert < d $. After that we are left to prove that the expression
in the parentheses tends weakly to zero as  $ \varepsilon \to 0$.
This proves assertion (iii) of the theorem.

\medskip \noindent Let us prove  assertion (iv), according to which the
solutions $\Psi_\pm(\x,\kk)$ are the kernels of the wave operators
$ \Omega _\pm$, whose existence and completeness are proved in
Theorem  \ref{t32} (see also \cite{JM4} for similar results). We
will base the proof on the formula (see \cite{Pe}):
\[\Omega_\pm \Phi= \mathrm{s}-\lim_{\varepsilon \to \mp 0 }
\int_{- \infty} ^{\infty} G(E+i \varepsilon)\mathcal{E}_0
(dE) \Phi,
\]
where $\mathcal{E}_0$ is the resolution  of identity of the
Laplacian $H_0$ of (\ref{lap}), and $G(z)=(H-z)^{-1}$. In the
$(\x, \kk)$ representation, usual in the scattering theory, this
formula can be written as follows:
\begin{equation}\label{OOex}
(\Omega_\pm \Phi)(\x)= \mathrm{s}-\lim_{\varepsilon \to \mp 0}
(\Omega_ \varepsilon \Phi)(\x),
\end{equation}
where
\begin{equation}\label{OeP}
 (\Omega_ \varepsilon \Phi)(\x)=
\int_{\TT^d}\Psi_{E_d(\kk)+i\varepsilon}(\x,\kk)
\hat{\Phi}(\kk) d \kk,
\end{equation}
and $\Psi_{z}(\x,\kk)$ is defined in (\ref{pSo}).

\medskip \noindent According to general principles, it suffices to prove
(\ref{OOex}) for a dense set of vectors of $L^2(\TT^d)$. We choose
a  set of  functions of the form
$\hat{\Phi}((k_1,k_2))=\hat{\Phi}_1(k_1)\hat{\Phi}_2(k_2)$, where
$\hat{\Phi}_{1,2}$ are  smooth and the support of
$\hat{\Phi}_1$ does not contain the  critical points of
$E_{d_1}$. Denoting the r.h.s. of (\ref{OeP}) by
$\Phi_{\varepsilon}(\x)$,  we  have to prove the relations:
\begin{enumerate}
\item[(a)]   $\lim_{\varepsilon \to \mp 0}\Phi_{\varepsilon}(\x)
=\Phi_{\pm}(\x) $;
\item[(b)]  $\lim_{\varepsilon \to \mp 0} \sum_{\x \in \ZZ^d} \vert
\Phi_\varepsilon(\x)-  \Phi_\pm(\x)  \vert^2=0.$
\end{enumerate}
\medskip \noindent where $\Phi_{\pm}(\x)$  are defined in (\ref{psifi}).
Both facts are proved in the Lemma \ref{l31}
below. Theorem \ref{t33} is proved.
%%%%%%%%%%%%%%%%%%%%%%%%%%%%%%%%%%%%%rem%%%%%%%%%%%%%%%%%%%%%%%%%%%%%%%%%%%%

\medskip \noindent {\it Remarks.} 1). Functions  $ \Psi_\pm(\x, \kk)$ are
analogs of the Sommerfeld solutions, which appear in the scattering
theory for potentials decaying in all directions and which
provide a complete set of generalized eigenfunctions for the part
of the spectrum that coincides with the spectrum of the Laplacian
\cite{Pe,Si}. Likewise, (\ref{LS}) is an analogue of the
Lippmann-Schwinger equation of scattering theory.

\smallskip
\noindent 2). According to formula (\ref{Som}), $
\Psi_\pm(\x,\kk)$ depends on the component $x_2$ of $
\x=(x_1,x_2), \; x_1 \in \ZZ^{d_1}, \; x_2 \in \ZZ^{d_2}$ via the
product of $e^{i k_2 \cdot x_2}$ and of a 1-periodic function of
the argument $\alpha \cdot x_2$, i.e. of a quasi periodic function
of $ x_2 \in \ZZ^{d_2}$ (recall that we assume in this section
that the components of the vector $\alpha$ in (\ref{vt}) are
rationally independent). This fact is in agreement with the widely
accepted idea, according to which generalized eigenfunctions of
absolutely continuous spectrum of differential and finite
difference operators with almost periodic coefficients have the
"almost Bloch" form, i.e the form of the product of a plane wave
and an almost periodic function with the same frequencies as the
coefficients (see e.g \cite{PF}).

\smallskip \noindent
3). According to formula (\ref{Go}), if $\vert E \vert > \nu$,
 the Green function $
G_0^{(\nu)}(\x;E+ i0)$ of the $\nu$-dimensional Laplacian decays
exponentially and if $\vert E \vert < \nu$  it decays as  $1/\vert x
\vert ^{\frac{\nu-1}{2}}$ for $\nu \geq
2$ (in the one dimensional case for $\vert
E \vert < 1$, $ G_0^{(1)}(x; E+ i0)$ behaves as $ e^{i \eta(E)|x|}$,
  where $\eta(E)$ is a
real valued function, see formula  (\ref{G01}) and (\ref{eta})
below ). As m varies the expression $E_d(\kk)-E_{d_2}(k_2-m\alpha)$
has values inside  $(-d_1,d_1)$ as well as
outside this interval, then the Green function
\[
G_0^{(d_1)}(x_1; E_d(\kk)-E_{d_2}(k_2-m\alpha)),
\]
entering the expression (\ref{Som}), may be exponentially decaying
or slowly decaying (i.e. as $1/|x|^{\frac{\nu-1}{2}}$). In other
words we can write, say for $\Psi_-$:
\begin{equation}\label{v-s}
 \Psi(\x,\kk) =  e^{2i\pi\kk\cdot\x} +
\Psi_{vol}(\x,\kk) + \Psi_{surf}(\x,\kk),
\end{equation}
where $\Psi_{vol}$ is the part of the sum in (\ref{Som}),
containing only slow decaying terms, and $\Psi_{surf}$ is the
part, containing the exponentially decaying terms.

\medskip \noindent
Recall now the definition of the surface states
according to \cite{JMP}
 (for other definitions see \cite{DS},\cite{JL3},\cite{JM4}).
%%%%%%%%%%%%%%%%%%%%%%%%%%%%%%%%%%%%%%%%%%%%%%%%%%%%%%%%%%%%%%%%%%%%%%%%%%%%%%%%%
%%%%%%%%%%%%%%%%%%%def%%%%%%%%%%%%%%%%%%%%%%%%%%%%%%%%%%%%%%%%%%%%%%%%%%%%%%%%%%%

\begin{definition}\label{d31}
Let  $\Psi_E$ be a generalized eigenfunction $\Psi_E$, corresponding to a
point $E$ of the spectrum of the operator $H$ of
(\ref{ham}) - (\ref{Vt}). We say that $\Psi_E$
is  a surface state, if for any $\varepsilon >0$ we have
\begin{equation}\label{ss}
\sup_{x_2 \in \mathbb{Z}^{d_2}} (1+|x_2|^{d_2/2+\varepsilon})^{-1}
\sum_{x_1 \in \mathbb{Z}^{d_1}} |\Psi_E\big((x_1,x_2)\big)|^2 < \infty .
\end{equation}
\end{definition}
\noindent Since the part $ e^{2i\pi\kk\cdot\x} +
\Psi_{vol}(\x,\kk)$ of the solution (\ref{v-s}) $ \Psi(\x,\kk)$ is
not decaying in the $x_1$-direction, the solution is not a "surface" state
but a "volume" state. Hence, we can say that Theorem \ref{t33} above
implies the existence of the volume states for all $E \in (-d,d)$.
Theorem \ref{t34} below implies that these generalized
eigenfunctions are complete in the interval $(-d,d)$. We conclude
that there is no surface states in the interval $(-d,d)$ of the
spectrum of the operator $H$  in the considered case of quasi
periodic surface potential (\ref{Vt}) - (\ref{vt}).
However, despite that   surface states are absent, the volume
states (\ref{v-s}) contain both a term, $e^{2i\pi \kk \cdot
\x}+\Psi^{vol}(\x,\kk)$ which slowly decays or even only
oscillates in $\vert x_1\vert$ , and a term,$\Psi_{surf}(\x,\kk)$,
 which exponentially decays in  $\vert x_1\vert$. They  are
respectively  the superposition of
reflected or transmitted waves which propagate inside the bulk and
of waves which propagate only along the subspace $\ZZ^{d_2}$.

\smallskip \noindent 4).
The scattering interpretation  (\ref{v-s}) of generalized
eigenfunction  (\ref{Som}) allows us to introduce transmission and
reflection amplitudes and coefficients (the latter as  square of
the modulus of the former). Consider the simplest case of $d_1=1$
and recall that:
\begin{equation}\label{G01}
 G_0^{(1)}(x;z)=  \frac{ie^{i\eta(z) |x|}}{ \sin \eta(z)}
 =-\frac{e^{i\eta(z) |x|}}{\sqrt{z^2-1}},
\end{equation}
where $-\cos \eta=z $,  or
\begin{equation}\label{eta}
 \eta(z ) =-i\log(-z+ \sqrt{z^2-1}),
\end{equation}
and we use the branch of the logarithm that has the cut along the
negative semi-axis and the branch of $\sqrt{z^2-1}$ fixed by the
condition $\sqrt{z^2-1}=z(1+ O(z^{-1})), z \to \infty $. In
particular $ \Im \eta(z) \geq 0$ for $\Im z \geq 0$ and
\begin{equation}
\eta( E+i0) \in  \left \{
\begin{array} {l} \label{eta1}
(0,\pi), \ \vert E \vert <1, \\
\pi  +i\mathbb{R}_+, \  E >1,
 \\ +i\mathbb{R}_+, \ E <-1.
\end{array}
\right.
\end{equation}
Combining these formulas and (\ref{Som}), we can present
$\Psi_{vol}(\x,\kk)$ in (\ref{v-s}) for $d_1=1$ as
\begin{equation}\label{Psiv}
  \Psi_{vol}(\x,\kk)=\sum_{m} \Psi_{m}(\kk)  e^{i
\eta_m(\kk)|x_1| +2i\pi( k_2 - \alpha m)\cdot x_2},
\end{equation}
 where $\sum_{m}$ denotes the sum of those terms in (\ref{Som}) for which
$\eta_m(\kk) := \eta(\lambda_m(\kk)+i0)$ is real, and
$\lambda_m(\kk)$ is defined by the equation: $\lambda_m(\kk)=
E_d(\kk)- E_{d-1}(k_2 - m\alpha)$. Recall that in the
one-dimensional scattering problem for the potential $v\delta(x),
x \in \ZZ$, the Sommerfeld solutions are (cf (\ref{Gpo}), and
(\ref{Tpo})):
 $$
\Psi_-(x,k) = e^{2i\pi k x} - \frac{iv}{iv+ \sin 2\pi k}
 e^{2i \pi \eta_-(k) \vert x \vert },
$$
where $\eta_-(k)=\eta(\cos2\pi k+i0), k \in \TT$.  Hence in this
case
 $$t(k)=  \frac{\sin2\pi k}{iv+
\sin 2\pi k}, \quad r_-(k)=-\frac{iv}{iv+ \sin 2\pi k} $$ are the
transmission and the reflection amplitudes. This
makes natural to view $$ t_0=1+\Psi_{0}(\kk), \quad r_0=\Psi_{0}(\kk), $$ where
$\Psi_{0}(\kk)$ is given by (\ref{Psiv}), as the transmission and the
reflection amplitudes of the plane waves scattered by the surface
potential (\ref{Vt}) and propagating in direction
$\kk$ of the incident wave and in the opposite direction. Likewise
it is natural to
view the coefficients $\Psi_{m}(\kk), m \geq 1$ of (\ref{Psiv}) as the
transmission and the reflection amplitudes of the scattered plane
waves  propagating in the directions $ (\eta_m,k_2 + m\alpha )$
and $ (-\eta_m,k_2 + m\alpha )$ respectively to the right and to
the left of the plane $x_1=0$. This scattering theory
interpretation of the solutions (\ref{Som}) is in agreement with
the form of the scattering matrix $\mathcal{S}$ in our case. We
use the general formula (see \cite{Pe}, formula (4.2.30)):
\begin{equation}\label{ST}
\mathcal{S}=1-\mathcal{T}, \quad \mathcal{T} = (-2i \pi) \;
\mathrm{s}-\lim_{ \stackrel {\varepsilon_1 \to 0} {\varepsilon_2
\to 0}} \int \delta_{\varepsilon_2}(H_0-\lambda) T(\lambda +i
\varepsilon_1) \mathcal{E}_0(dE),
\end{equation}
where $ \delta_{\varepsilon}(A)=(2i \pi)^{-1}
[(A+i \varepsilon)^{-1} -(A- i\varepsilon)^{-1}] $,
 $ T(z)$ is defined in
(\ref{GT}), $ \mathcal{E}_0 $ is the resolution of identity of $
H_0$, and the limits have to be carried out in the following
order: first $\varepsilon_1 \to 0$, second $\varepsilon_2 \to 0$.
Formula (\ref{ST}) implies that for any sufficiently smooth
function $\hat{f}$ on $\mathbb{T}^d$  we have:
$$
(\mathcal{T}\hat f)(\kk)= (-2i \pi) \; \mathrm{s}-\lim_{ \stackrel
{\varepsilon_1 \to 0} {\varepsilon_2 \to 0}} \int_{T^d}
\delta_{\varepsilon_2}(E_d(\kk)- E_d(\p)) T(\kk,\p;E_d(\p)+i
\varepsilon_1) {\hat f}(\p)d\p.
$$
By using formula  (\ref{Tkp})
for the kernel of the $T$-operator, it can be shown that the
generalized kernel $ \mathcal{T}(\kk,\p)$ of the
$\mathcal{T}$-matrix of (\ref{ST}) is:
\begin{eqnarray}\label{kerT}
 \mathcal{T}(\kk,\p) &=& -2i \pi \delta(E_d(\kk)- E_d(\p)) T(\kk,
\p;E_d(\p)+i0)\\ & =& -2i \pi \delta(E_d(\kk)- E_d(\p)) \sum_{m=0}
^\infty t_m(k_2 ;E_d(\kk)+i0) \delta(k_2 + m \alpha -p_2).
\nonumber
\end{eqnarray}

\medskip \noindent Now we formulate and prove the lemma that
was used in the proofs of assertions (iii) and (iv) of Theorem \ref{t33}.
%%%%%%%%%%%%%%%%%%%%%%%%%%%%%%%lemma31%%%%%%%%%%%%%%%%%%%%%%%%%%%%%%%%%%%%%
\begin{lemma}\label{l31}
 Let  $ {\hat \Phi}_{1,2}: \TT^{d_{1,2}} \to \CC $ be smooth
 functions. Assume that the support of $\hat \Phi_{1}$
  does not contain the critical points of $E_{d_1}$:
\begin{equation}\label{phi1}
\rm{supp} \; {\hat \Phi}_1 \bigcap \{ k_1 \in \TT^{d_1}: \nabla_1
E_{d_1}(k_1) =0\}= \emptyset.
\end{equation}
Set for $ \varepsilon \not= 0$: $$ \Phi_\varepsilon(\x)=
\int_{\TT^d} \Psi_{E_d(\kk)+i\varepsilon}(\x,\kk) {\hat
\Phi}(\kk)d\kk,$$ where $\Psi_z(\x,\kk)$ is defined in (\ref{Gk}),
(\ref{pSo}), and in (\ref{psiz}), and ${\hat \Phi}(\kk)$ is of the
form $ {\hat \Phi}((k_1,k_2))= {\hat \Phi}(k_1) {\hat \Phi}(k_2)$.
Then:
\begin{equation}\label{pep}
\lim_{\varepsilon \to \mp 0} \sum_{\x \in \ZZ^d} \big|
\Phi_\varepsilon(\x) - \Phi_\pm(\x)\big|^2=0,
\end{equation}
where $ \Phi_\pm(\x,\kk)$ are defined in (\ref{lim}) and in
(\ref{psifi}).
\end{lemma}
%%%%%%%%%%%%%%%%%%%%%%%%%%%%FIN du l31%%%%%%%%%%%%%%%%%%%%%%%%%%%%%%%%%%%%%
\noindent {\it Proof.} By using (\ref{psiz}), we find that for any
$\varepsilon \not=0$:
\begin{eqnarray}\label{pe}
 \Phi_\varepsilon (\x) & = &  \Phi( \x) +
\sum_{m=0}^\infty \int_{\TT^{d_1}} dp_1{\hat   \Phi}_1(p_1)
\int_{\TT^{d_2}}
 dp_2 e^{2i\pi(k_2- m\alpha)\cdot x_2}{\hat \Phi}_2(p_2)
  \\ &  \times&
 t_m(k_2, E_d(\kk)+i\varepsilon)G_0^{(d_1)}
 (x_1, E_d(\kk)- E_{d_2} (k_2 -m\alpha) +i\varepsilon),
 \nonumber
\end{eqnarray}
where $\Phi$ is the Fourier transform of $\hat \Phi$. The
integrals and the series in this formula converge and can be
written in any order because of the bound (\ref{bou}) applicable
in view of  (\ref{phi1}). The integral representation (\ref{Go})
for $ G_0^{(d_1)}$ allows us to rewrite the last formula as
follows:
\begin{equation}\label{pFe}
 \Phi_\varepsilon(\x)=
\Phi(\x)+ \int_{\TT^d}e^{2i\pi \kk \cdot \x} \hat{\Psi}_
\varepsilon (\kk)d\kk,
\end{equation}
where
\begin{eqnarray}\label{Fe}
\hat{\Psi}_\varepsilon (\kk)&=& \sum_{m=0}^\infty \int_{\TT^{d_1}}
dp_1 \frac{{\hat   \Phi}_1(p_1) {\hat \Phi}_2(k_2 +m\alpha)}
{E_d(\kk)-  E_{d_1}(p_1) -E_{d_2} (k_2 + m\alpha) +i\varepsilon}\\
&\times & t_m(k_2,E_{d_1}(p_1) +E_{d_2} (k_2 + m\alpha)
+i\varepsilon). \nonumber
\end{eqnarray}
This series and the integral are convergent because the modulus of
the denominator is bounded from below for $\varepsilon \not=0$,
and because of bound (\ref{bou}).

\medskip \noindent Now  we will prove that for
any $\kk \in \TT^{d}$, the limits $
\lim_{\varepsilon \to \mp 0} \hat{\Psi}_\varepsilon (\kk) \equiv
\hat{\Psi}_\mp (\kk)$ exist and that the convergence is bounded.
Consider the case $\hat{\Psi}_-$
 for the  sake of definiteness. The building block of
the coefficient $ t_m(k_2, E+i\varepsilon)$
in (\ref{Fe}) is the function $ {\hat \gamma}_0 (k_2, E+i\varepsilon)=
 G_0^{(d_1)}(0, E -E_{d_2} (k_2) +i\varepsilon))$.
 This function is real analytic in $k_2 \in \dot \TT^{d_2}$
(see (\ref{Tdot}) for the definition of  $\dot \TT^{d_2}$), and
in $ E \in (-d+\gamma, d-\gamma)$ for any fixed (small)
$\gamma>0$, (see (\ref{Go}) and (\ref{Ek})). By using identity
(\ref{lz}) for $G_0^{(d_1)}$, we can write
the $m$th term of formula (\ref{Fe}) as:
\begin{eqnarray}\label{mth}
&&\int_0^\infty dt e^{-\varepsilon t-it (E_d(\kk) -E_{d_2} (k_2 +
m\alpha))} {\hat   \Phi}_2(k_2 + m\alpha)  \\ &&\times
\int_{\TT^{d_1}} dp_1 {\hat   \Phi}_1(p_1) e^{-it E_{d_1}(p_1)}
t_m(k_2,E_{d_1}(p_1) +E_{d_2} (k_2 + m\alpha) +i\varepsilon)).
\nonumber
\end{eqnarray}
Since the support of ${\hat   \Phi_1}$ does not contain critical
points of $E_{d_1}$ and since $G_m(k_2,E+i\varepsilon)$ is real
analytic in $ k_2 \in \TT^{d_2}$  and in $E \in (-d + \gamma, d-
\gamma), \gamma >0$
 for all $\varepsilon \geq 0 $, we can integrate by parts
 twice in components of $p_1 \in \TT^{d_1}$, and obtain  an expression
of the form $t^{-2}\Phi_m(\kk,p_1, \varepsilon)$, where $\Phi_m$
is bounded in $\kk \in \TT^{d}, \; p_1 \in \TT^{d_1}$ and
$\varepsilon>0$. This allows us to make the limit $ \varepsilon
\to +0$ in (\ref{mth}) and obtain a bounded in $\kk$  expression.

\medskip \noindent Besides, $\Phi_m$ is a linear combination of the first
and second partial derivatives in components of $ p_1 \in
\TT^{d_1}$ of the integrand in (\ref{mth}). The derivatives are
linear combination of products of bounded (and smooth) in $\kk \in
\TT^{d}, p_1 \in \TT^{d_1}$ for $ \varepsilon >0$, and independent
of $m$ functions, multiplied by the first and the second partial
 derivatives  in
components of $ p_1 \in \TT^{d_1}$
 of $ \prod_{l=1}^m {\hat b}(k_2 +l \alpha, E_{d_1}(p_1) +
 E_{d_2} (k_2 + m\alpha) +i\varepsilon))$.
This leads to the bound $ \vert \Phi_m(\kk,p_1, \varepsilon)\vert
\leq c_1m^2e^{ - c_2m}$
 where $c_1,c_2 >0 $ and are independent of $m,\kk,p_1$
 and $\varepsilon$.
The  bound allows us to make the limit $\varepsilon \to+0 $ in
(\ref{Fe}) for any $\kk \in \TT^{d}$:
\begin{equation*}
%\label{FeF}
\lim_{\varepsilon \to +0} \hat \Psi_\varepsilon (\kk)=\hat \Psi_-(\kk),
\end{equation*}
and to obtain the bound $ \vert \Psi_\varepsilon (\kk)\vert \leq
\mathrm{const}$,
 valid for any $ \varepsilon \geq 0$ and $\kk \in \TT^{d}$.
 Now the Lebesgue dominated
convergence theorem and relation (\ref{lim}) proved above lead
to the representation:
\begin{equation}\label{FeF}
 \Psi_-(\x)=  \Phi(\x)+ \int_{\TT^d}e^{2i\pi\kk.\x}
 \hat \Psi_-(\kk)d\kk.
\end{equation}
Subtracting this relation from  (\ref{pFe}) and
applying to the  result the Parseval equality, we
obtain that:
 \begin{equation}
  \sum_{\x \in \ZZ^d} \vert \Psi_\varepsilon (\x)-\Psi_-
  (\x)
\vert^2 =   \int_{\TT^d} \vert   \hat \Psi_\varepsilon(\kk)
-  \hat \Psi_-(\kk) \vert^2 d\kk.
\end{equation}
Thus (\ref{FeF} and the Lebesgue theorem imply (\ref{pep}). Lemma
is proved.
%%%%%%%%%%%%%%%%%%%%%%%theorem34%%%%%%%%%%%%%%%%%%%%%%%%%%%%%%%%%%%%%%

\begin{theorem}\label{t34}
 Let  $H=H_0+V$  be the self-adjoint operator on $l^2(\ZZ^d)$, defined by
(\ref{ham}) - (\ref{par}) in which the vector $\alpha \in
\R^{d_2}$ has rationally independent components. Then the family
$\{ \Psi_{z}(\x,\kk); \x \in \ZZ^d\}_{\kk \in {\dot \TT}^d}$,
defined in Theorem \ref{t33} (see (\ref{Gk}), (\ref{lim}), and
(\ref{Som})), is the complete system of generalized eigenfunctions
of $H$ in the part $(-d,d)$ of the spectrum of $H$, i.e.:
\begin{itemize}
\item[(i)] for any $f \in l^2(\ZZ^d)$, the series:
\begin{equation}\label{FPsi}
F_\pm(\kk)= \sum_{\x \in \ZZ^d} \overline{ \Psi_\pm(\x,\kk)}f(\x)
\end{equation}
converges in $l^2(\ZZ^d)$;
\item[ (ii)] if $\mathcal{E}_H (\Delta)$ is the spectral projection of $H$,
corresponding to the closed interval $\Delta= [a,b] \subset
(-d,d)$, then
\begin{equation}
\Vert \mathcal{E}_H (\Delta)f\Vert^2= \int_{\kk \in \dot\TT^d: E_d(\kk)
\in \Delta} \vert
 F_\pm(\kk) \vert^2d \kk; \label{com}
\end{equation}
where $E_d(\kk)$ is defined in  (\ref{Ek}));
\item[ (iii)] the following relation is valid
\begin{equation}
\Vert H\mathcal{E}_H (\Delta)f\Vert^2= \int_{\kk \in \dot\TT^d:
E_d(\kk) \in \Delta} \vert
 E_d(\kk)F_\pm(\kk) \vert^2d\kk.
 \label{HDel}
\end{equation}
\end{itemize}
\end{theorem}
%%%%%%%%%%%%%%%%%%%%%%%FIN du t34%%%%%%%%%%%%%%%%%%%%%%%%%%%%%%%%%%%
\noindent{\it Proof.} We write  the Hilbert identity
for the Green function
 $G(\x,\y ; z_{1,2} ), \  \Im z_{1,2}\not=0 $:
\begin{equation}\label{Hil}
 G(\x,\y,z_1) -  G(\x,\y,z_2)= (z_1-z_2) \sum_{ \s \in \ZZ^d}
G(\x,\s; z_1) \overline{G(\y,\s; { \bar z}_2)}.
\end{equation}
By using the Parseval equality for the Fourier transform with
respect to the variable $\s$ in the r.h.s. of this identity, we rewrite it
as follows: $$ \int_{ \kk \in \TT^d} d\kk
G(\x,\kk,z_1)\overline {G (\y,\kk, { \bar z}_2)},$$ where $G(\x,\kk,z)$
is the Fourier transform of $ G(\x,\y,z)$ in the
 second variable $ \y$, defined in (\ref{Gk}).
Multiply now resulting relation by $\overline{f(\x)}f(\y) $, where
$f$ has compact support in $ \ZZ^d$ and sum over $\x,\y \in
\ZZ^d$. This yields: $$((G(z_1)- G(z_2))f,f)= \int_{\dot \TT^d}
d\kk \frac{z_1-z_2}{(E_d(\kk)-z_1)(E_d(\kk)-z_2)}\overline{ F_{z_1}(\kk)}
{F}_{\overline{z}_2}(\kk),$$ where
\begin{equation}\label{hz}
F_z(\kk)= \sum_{\x \in \ZZ^d} \overline {\Psi_z(\x,\kk)} f(\x),
\end{equation}
and $\Psi_z(\x,\kk)$ is defined in (\ref{pSo}). Setting $z_1={\bar
z}_2=E+i\varepsilon,  \varepsilon>0$, we get:
\begin{equation}\label{GzF}
\frac{1} {\pi} \Im(G(E+i \varepsilon)f,f)= \frac{1}{\pi} \int_{ \dot
\TT^d} d\kk \frac{\varepsilon}{(E_d(\kk)-E)^2+ \varepsilon^2}
 \vert F_{E+i\varepsilon}(\kk) \vert^2.
\end{equation}
$ \Delta= [a,b] \in (-d,d)$, we obtain in the l.h.s. of the
resulting relation the expression $\Vert \mathcal{E}_H
(\Delta)f\Vert^2$.  can be
continued in $z$ to the real $ z=E_d(\kk)+i0 \in \Delta$, and
that the continued function is uniformly continuous
in $\kk \in \{ \kk\in\dot \TT^d:E_d(\kk)\in\Delta \}$,
where $\dot \TT^d$ is defined in (\ref{Tdot}). Since $f$ is of
compact  support in $\ZZ^d$, it suffices to show that
 $\Psi_z(\x,\kk)$ possess this
property for any fixed $\x \in \ZZ^{d}$. But this fact is proved
Theorem \ref{t33}. Thus we have established (ii) for the case
where $f$'s  of finite support. The extension to $f$'s belonging
to $l^2(\ZZ^d)$ is based on the standard arguments of spectral
theory (see e.g. \cite{Si,RS}). This proves assertions (i) and
(ii). As for assertion (iii), it follows from (ii) and from the
spectral theorem.

%%%%%%%%%%%%%%%%%%%%%%%%%%%%%%%%%%%%%%%%%%%%%%%%%%%%%%%%%%%%%%%%

%%%%%%%%%%%%%%%%%%%%%%%%%%%%%%%%%SECTION4%%%%%%%%%%%%%%%%%%%%

\section{ The  Periodic Case}

\setcounter{equation}{0}
In this section we consider the operator $H=H_0+V$ of (\ref{ham})
- (\ref{par}) in which $d_1=d_2= 1$ and $\alpha$ is a rational
number: $\alpha = p / q, \; p \in \ZZ , q \in \ZZ \setminus \{0\}$,
i.e. for periodic potentials $v$ of (\ref{vt}). We show that in
this case the whole spectrum of $H$ is absolutely continuous and
we construct corresponding generalized  eigenfunctions. It turns out that
there are two types of generalized eigenfunctions. Both types have the
Bloch-Floquet form in the longitudinal coordinates
$x_2$ but behave differently in the transverse coordinate $x_1$.

\medskip \noindent We will follow the same  strategy as in the preceding
section namely the construction of  generalized eigenfunctions
based  on the formulas for the Green function of Section 2 and on
formulas (\ref{Gk}), (\ref{pSo}) and (\ref{lim}) of Theorem
\ref{t33}. Thus we have to analyze the behavior of the Green
function as the spectral parameter tends to the real axis. Our
first goal is to find the set of energies for which the
limit $G(\x,\y, E+i0)$ exists and is  bounded, i.e. the
purely absolutely continuous part of the spectrum. We shall see that
unlike the quasiperiodic case, where this set is $[-d,d]$, in the
periodic case the whole spectrum is pure absolutely continuous.
The spectrum which lies outside $[-d,d]$  consists of
surface states only. As for the  part in the interior of
 $[-d,d]$, it  consists of the  volume states whose energies
occupy the whole interval $[-d,d]$, and  of the
surface states that may exist under certain conditions.

\medskip \noindent For any $ z \in \CC, \Im z \not=0$, and $m=1,...,q$
define the function:
\begin{equation}\label{Pm}
  P_m(k_2;z)=\sigma^m \prod_{l=1}^m {\hat b}(k_2  +l\alpha; z ),
  \forall k_2 \in \TT,
\end{equation}
where $\sigma$ and ${\hat b}$ are defined  by (\ref{sig}),
(\ref{Ga0b}),(\ref{bGak}) and $\TT=(0,1]$. Then, by using Lemma
\ref{l22}, we obtain for $\alpha=p/q$:
\begin{equation}\label{Pq}
\widehat {((bu)^q \varphi)} (k_2)=
 \sigma^q\prod_{l=1}^{q} {\hat b}(k_2  +l\alpha; z){\hat \varphi}(k_2)=
P_q(k_2;z){\hat \varphi}(k_2),
\end{equation}
where the operator $u$ is defined in (\ref{u}). We conclude that
$(bu)^q$ is a multiplication operator  by the function
$P_q$ in the space $L^2(\TT)$.
%%%%%%%%%%%%%%%%%%%%%%%%%%%%%%%%%%%%%%%%%%THEOREM41%%%%%%%%%%%%%%%%%%%%%%%%=
%%%%%%%
\begin{theorem}\label{t41}
 Let $H=H_0 + V$ be the operator defined by (\ref{ham}) -
 (\ref{par}) in which  $d_1=d_2= 1$ and
 $\alpha = p / q, p \in \ZZ, q \in
\ZZ\setminus  \{0 \}$ is a rational parameter. Then the Green
function $G(\x,\y;z)= (H-z)^{-1}(\x,\y), \ \x,\y \in \ZZ^2$ of $H$
can be written in the form:
\begin{eqnarray}\label{GFp}
G(\x, \y;z)&=&G_0^{(2)}(\x-\y;z) + \sum_{m= 0}^{q}\int_{\TT} dk_2
e^{2i\pi k_2 (x_2-y_2)}t_m(k_2;z)
\\ &\times&  G_0^{(1)}(x_1;z + \cos 2 \pi k_2) G_0^{(1)}\big(y_1;z+ \cos 2
\pi (k_2+m\alpha)\big)e^{-2i\pi m\alpha  y_2}, \nonumber
\end{eqnarray}
where
\begin {eqnarray} \label{tmp}
t_m(k_2;z)&=&  \displaystyle{\frac{g}{g \hat \gamma_0(k_2;z) +i}}
\\ &\times& \left\{
\begin{array}{ll} \nonumber
-1,\ &  m=0; \\
\displaystyle{\frac{1}{1-P_q(k_2;z)}
\frac{2i\sigma}{g \hat \gamma_0(k_2 +\alpha;z)+i}}, & \ m=1; \\
\displaystyle{\frac{1}{1-P_q(k_2;z)}\frac{2i\sigma}{g \hat
\gamma_0(k_2+m\alpha;z)+i}}P_{m-1} (k_2;z), & \ m \geq 2,
\end{array}
\right.
\end{eqnarray}
$ G_0^{(\nu)}(.;z), \; \nu=1,2$ is the Green function (\ref{G01})
of the $\nu$-dimensional discrete  Laplacian, $\hat
\gamma_0(.;z)$, and $\hat b(.;z)$ are defined in (\ref{GaxGak})
and in (\ref{bGak}).
\end{theorem}
%%%%%%%%%%%%%%%%%%%%%%%%%%FINTHM%%%%%%%%%%%%%%%%%%%%%

\noindent The proof of the theorem is based on the same argument
as that used in the proof of Theorem  \ref{t21}.

\medskip \noindent Formulas  (\ref{GFp}) and (\ref{tmp}) suggest that the
spectrum $\sigma(H)$ of $H$  contains the  set $S=\{E \in \R:
\exists k_2 \in \TT  ;  P_q(k_2,E)=1\}$. We prove below that indeed,
the limit
$G(\x,\y,E+i0)$ exists and is bounded for all $E \in S \setminus  D $
where $D$ is a discrete set.

\medskip \noindent For any  $1>\gamma>0, \; E \in
\mathbb{R}$ and $l=1,..., q$, define the sets:
\begin{equation}\label{Kgal}
K_\gamma^l (E)=\{k_2 \in \TT:
 \; E+ \cos 2 \pi (k_2 +  l\alpha) \in  [-1 +\gamma,1 -\gamma ]\},
\end{equation}
and
\begin{equation}\label{Kc}
K_\gamma (E)=\bigcup_{l=1}^{q} K_\gamma^l (E), \quad
K_\gamma^{c}(E)=\TT \setminus  K_\gamma (E).
\end{equation}
It follows from formula  (\ref{Pq}), Lemma \ref{lA3}, and from the argument of the proof of Theorem \ref{t31},
that for any $0<\gamma <1$ there exists $\delta(\gamma)>0$ such
that the inequality $ {\sup}_{\varepsilon>0} \vert P_q(k_2, E
+i\varepsilon)\vert < 1-\delta $ is valid uniformly in $ k_2 \in
K_\gamma (E)$. This means that the function $(1-P_q(k_2,E+i0))
^{-1}$ is well defined  and bounded on the sets,
 \[
 K(E)=\bigcup_{0<\gamma<1} K_\gamma (E)
 \]
and that possible singularities of this function which are given
by the "band-equation":
\begin{equation}\label{be}
P_q(k_2,E)= 1,
\end{equation}
where $P_q(k_2,E)=P_q(k_2,E+i0)$, are localized on
$K^{c}(E)$. It is natural to think that
energies, satisfying the band equation (\ref{be}) for some $k_2 \in
\TT$ belong to the spectrum of $H$. The following proposition
describes properties of solutions of the band equation.

%%%%%%%%%%%%%%%%%%%%%%%%%%PROP41%%%%%%%%%%%%%%%%%%%%%%%%%%%%%%%%%%%
\begin{proposition}
\label{l41} For any $2\leq q<\infty $ the band equation
(\ref{be}) admits a finite number $N_{q}^{\prime }$ of positive
solutions  $0\leq E_{1}(k_{2})<...<E_{N_{q}^{\prime }}(k_{2})<
\infty $  (the positive energy band functions), and a finite
number $N_{q}^{\prime \prime }$ of negative solutions $-\infty <
E_{-N_{q}^{\prime \prime }}(k_{2})<...<E_{-1}(k_{2})\leq 0$ (the
negative energy band functions).

\medskip \noindent
The functions $E_{j},\; j=-N_{q}^{\prime \prime
},...-1,1,...N_{q}^{\prime }$ are $1/q$-periodic in $k_{2}$
, and are real analytic in the interior
of their respective domains $\mathcal{D}_{j}\subset \TT$ (each domain $%
\mathcal{D}_{j}$ is a  closed  subset of
$\mathbb{T}$).

\medskip \noindent
Moreover, the band functions are separated in the sense that:
\begin{enumerate}
\item[(i)]  for any $j=-N_{q}^{\prime \prime
},...-1,1,...N_{q}^{\prime }$ there exists a finite subset
$\mathcal{D'}_{j}$
of $\mathcal{D}_{j}$, such that for all $\mathbf{k}\in \TT\times (%
\mathcal{D}_{j}\setminus \mathcal{D'}_{j})$ we have:
\begin{equation}
|E_{j}(k_{2})-E_{2}(\mathbf{k})|>0,  \label{sv}
\end{equation}
where $E_{2}(\mathbf{k})=-\cos 2\pi k_1 -\cos 2\pi k_2$;
\item[(ii)] there exists a positive constant $\eta _{q}>0$ such
that for any $j,j^{\prime }=-N_{q}^{\prime \prime
},...-1,1,...N_{q}^{\prime },\;j\not=j^{\prime }$ we have:
\begin{equation}
\inf_{k_{2}\in \mathcal{D}_{j}\bigcap
\mathcal{D}_{j^{\prime}}}|E_{j}(k_{2})-
E_{j^{\prime }}(k_{2})|\geq \eta _{q}>0.
\label{sb}
\end{equation}
\end{enumerate}
\end{proposition}
%%%%%%%%%%%%%%%%%%%%%%%%%%%%%%%FINPprop%%%%%%%%%%%%%%%%%%%%%%
\noindent The proof of the proposition will be given after the proof of
Theorem \ref{t46}.
%%%%%%%%%%%%%%%%%%%%%%%%%%FIN-lmm%%%%%%%%%%%%%%%%%%%%%

\medskip \noindent
The band function $E_j, \; j=-N_{q}^{\prime
\prime },...-1,1,...N_{q}^{\prime }$ defined in Proposition (\ref{l41}) determine the band-gap
structure of the spectrum of  the periodic in $x_2$ operator $H$
in the following sense,

%%%%%%%%%%%%%%%%%%%%%%%%%%%%%%%%%%%%%%%%%THEOREM41%%%%%%%%%%%%%%%%%%%%%%%%%=
%%%%%%
\begin{theorem}\label{t42}
 Let  $H=H_0+V$ be  the operator defined  in Theorem
\ref{t41}. Then for all  rational parameter $
 \alpha , \; g \neq 0 $, and $ \omega \in [0,1]$ the  spectrum $\sigma(H)$
 of $H$ is a finite union of closed intervals (energy bands):
\begin{equation}\label{sigp}
\sigma(H)= \bigcup_{j=-N''_q}^{N'_q}  \mathrm{Ran} \; E_j \cup
[-2,2].
\end{equation}
\end{theorem}
%%%%%%%%%%%%%%%%%%%%%%%%%%FIN-THM%%%%%%%%%%%%%%%%%%%%%
\noindent The assertion  that $(-d,d) $ is in the spectrum of $H$ is
 a consequence of Theorem \ref{t22},  the rest of the theorem will be
proved after the proof of Theorem \ref{t45}.

\medskip \noindent Let us  now  define  the set $\mathcal{E}_c$ of critical
energies as
\[\mathcal{E}_c= \{ E\in \R: \exists j \in
\{-N_{q}^{\prime \prime },...,N_{q}^{\prime }\} ,\; \exists k_2 \in
\TT, \;  E_j(k_2) = E \; \mathrm{and \;  either \;}
\displaystyle{\frac{ dE_j}{dk_2}(k_2)} = 0 \; \mathrm{ or} \;
k_2 \in \partial \mathcal{D}_j \}.
\]
Denote
\begin{equation}
D= \mathcal{E}_c \bigcup \{-d,d\} \label{D},
\end{equation}
and notice that  because of Proposition \ref{l41} $D$ is a
discrete subset of $\R$.

%%%%%%%%%%%%%%%%THEOREM43%%%%%%%%%%%%%%%%%%%%%%%%%%%%%%%%%%%%%%%%%%%%%%%%%%=
%%%%%%%%%%%%%%%%%%%%%%%
\begin{theorem}\label{t43}
 Let   $H=H_0+V$ be the operator defined
 in Theorem  \ref{t41}, and let $G(\x,\y;z)$ be its Green function.
 Then for any  rational
 $\alpha$, $g \in
\R$, and $ \omega \in [0,1)$ the limit $G(\x, \y; E+i0)$ exists
and is bounded for any $E \in \sigma(H) \setminus D$ and $\x,\y
\in \ZZ^2$, where $D$ is defined in (\ref{D}). In particular the
spectrum of $H$ is  absolutely continuous.
\end{theorem}
%%%%%%%%%%%%%%%%%%%%%%%%%%FIN-THM43%%%%%%%%%%%%%%%%%%%%%
\noindent {\it Proof.} For any  $ E \in \sigma(H) \setminus D$ set
$z= E+i \varepsilon, \varepsilon>0$ and  fix $0< \gamma < 1$. By
using formula (\ref{GFp}) we can write that
\begin{equation}\label{dG}
G(\x, \y; z)=G_{1,\gamma}(\x, \y; z) + G_{2,\gamma} (\x, \y; z),
\quad \x, \y \in \ZZ^2,
\end{equation}
where
\begin{eqnarray}\label{G2}
 G_{1,\gamma}(\x, \y; z)&=&G_0^{(2)}(\x-\y;z) + \int_{ K_\gamma (E)} dk_2
e^{2i\pi k_2(x_2-y_2)}
\sum_{m=0}^{q} t_m(k_2;z)\\&\times&
G_0^{(1)}(x_1;z+ \cos 2 \pi
k_2) G_0^{(1)}(y_1;z+ \cos 2 \pi (k_2+m\alpha))e^{-2i\pi
m\alpha y_2},\nonumber
\end{eqnarray}
 $ K_\gamma (E)$ and $ t_m$ are  defined in (\ref{Kc}) and in (\ref{tmp}),
and
\begin{equation}\label{G1}
G_{2,\gamma}(\x, \y; z)=G(\x, \y; z) - G_{1,\gamma}(\x, \y; z).
\end{equation}
Since the inequality
$\sup_{\varepsilon \geq 0} \vert P_q(k_2, E +i\varepsilon)\vert <
1-\delta$ is valid uniformly on $ K_\gamma (E)$, the same arguments
as  in the proof of Theorem \ref{t31} imply that the limit
$G_{1,\gamma}(\x, \y; E+i0)$ exists  and is bounded.

\medskip \noindent Hence, to prove the theorem we have to
show the same property for
the term $ G_{2,\gamma}(\x, \y; z)$ of (\ref{dG}).
 We first note that by Proposition \ref{l41} this term can be
rewritten as
\begin{equation}\label{G21}
 G_{2,\gamma}(\x, \y; z)=\int_{ K^c_\gamma (E)} dk_2
\frac{ g_{2,\gamma}(\x, \y, k_2;z)}{1- P_q(k_2;z)},
\end{equation}
where for any $0< \gamma <1$, $\varepsilon \geq 0$ and  $(\x,\y)
\in \ZZ^2 \times \ZZ^2$, $ g_{2,\gamma}(\x, \y,\big. ;z)$ are smooth
functions  on $K_\gamma^c (E)$. Now in order to compute
the integral in the r.h.s. of (\ref{G21}), consider the level
sets:
$$ S_j= S_j(E,\gamma)=\{  k_2 \in K_\gamma^c (E): E_j(k_2)=
E\}, \quad j=-N''_q,..., N'_q,
$$
and the following neighborhoods
$\nu_j$ of $S_j$: $$ \nu_j=\nu_j(E,\gamma, \eta)= \{ k_2 \in
K_\gamma^c (E): \vert E_j(k_2)-E \vert \leq  \eta\}. $$
 If  $ \eta$ is  small enough, then  Proposition \ref{l41}
implies the relation:
$ \nu_j \cap \nu_{j'} = \emptyset $ if $j \not= j'$. Thus to prove
that $G_{2,\gamma}(\x, \y; E +i \varepsilon)$ exists and  is bounded as
$\varepsilon \to 0$,  it suffices to show that this holds for
\begin{equation}\label{G22}
 G_{2,\gamma,j}(\x, \y;  E +i \varepsilon)=\int_{ \nu_j} dk_2
\frac{ g_{2,\gamma}(\x, \y, k_2; E +i \varepsilon)}{1-
P_q(k_2;z)}, \quad  j=-N''_q,..., N'_q.
\end{equation}
Since $\eta$ is small enough and $E \not \in D$, we can
parameterize $\nu_j$ by  the  local coordinate $\tilde E$ defined
by the relation $\tilde E=E_j(k_2)$. Denoting $\varphi_j$ the
respective change of variables and $ J_{ \varphi_j}$  its
Jacobian, we have
 \begin{equation}\label{G23}
 G_{2,\gamma,j}(\x, \y; E +i \varepsilon)=
\int_{-\eta}^\eta d\tilde E \frac{ g_{2,\gamma}\circ \varphi_j(\tilde
E)}{1- P_q\circ \varphi_j(\tilde E)} J_{ \varphi_j}, \quad
j=-N''_q, \cdots, N'_q.
\end{equation}
Suppose now that $\eta$ and $\varepsilon$ are so small that we can
write:
\[ 1- P_q( \varphi_j(\tilde E), E +i
\varepsilon)= (\tilde E - E - i\varepsilon) p_j(\tilde E; E + i
\varepsilon), \tilde E \in [-\eta, \eta],
\]
where $ p_j, \ j=-N''_q, \cdots, N'_q$ are smooth and non
vanishing functions on the interval $[-\eta, \eta]$ such that
\[ \vert p_j (., E) \vert
\geq C \vert {\partial}_E P_q(., E) \vert + O(\eta)
+O(\varepsilon)
 \]
for some strictly positive constant $C$. Moreover it follows from
the proof of Proposition  \ref{l41} (see formula \ref{pder} ) that
 \[\vert {\partial }_E P_q(\varphi_j(\tilde E), E)  \vert \not=0,
 \quad \tilde E \in [-\eta, \eta].
 \]
 Then standard arguments imply the
existence and the boundedness of $ G_{2,\gamma,j}(\x, \y; E+i0)$,
$j=-N''_q, \cdots, N'_q$ hence the existence and the boundedness of $
G_{2,\gamma}(\x, \y;E+i0)$. The theorem is proved.

%%%%%%%%%%%%%%%%%%%%%%%%%%%%fin%%%%%%%%%%%%%%%

 %%%%%%%%%%%%%%%%%%%%%%%%%%%%%thmtwop%%%%%%%%%%%%%%%%%
 
\medskip \noindent
The  last theorem  together with the arguments of the proof
of Theorem  \ref{t32} lead to:

\begin{theorem}\label{twop}
 Under the conditions of the Theorem \ref{t41},
  the  wave operators $\Omega_\pm$ for the pair $(H,H_0)$ defined  in
(\ref{ham}) - (\ref{par}) with a rational $\alpha$  exist and are
complete for any closed interval $ \Delta=[a,b] \subset (-d,d)
\setminus \cup_{j=-N''_q}^{ N'_q} { \rm Ran} E_j  $.
\end{theorem}
%%%%%%%%%%%%%%%%%%%%%%%%%%%fin du twop%%%%%%%%%%%%%%%
\noindent Our next theorem  shows that surface states
 (see definition \ref{d31}) exist and are bounded.
They can be labelled  by the "quasi-momentum" $k_2 \in \TT /q$,
such that respective eigenvalues are given by the band functions:
$E=E_j(k_2)$. The "volume" states that do not belong to $l^2(\ZZ)$
in $x_1$ are labelled by the "momentum" $\kk \in \TT^2$, such that
the corresponding eigenvalues are given by the dispersion law of
the Laplacian: $E=E_2(\kk)= - (\cos2 \pi k_1 +\cos 2 \pi k_2)$. We
consider here only the non-degenerate case, i.e. the case where
chosen pairs ($k_2, E= E_j(k_2))$, and $(\kk,E= E_2 (\kk))$ are
such that $E_j(k_2)\not= E_{2}(\kk)$. By Proposition \ref{t41}
this property  is valid  for all energies except a finite set.

\medskip \noindent  Consider the set:
$$ \TT_j^2= \{ \kk=(k_1, k_2) \in
\dot\TT^2, \:  k_2 \in  \TT_j   \},  \;  j=-N''_{q},...,
N'_q, $$
where $\dot \TT^2$ is defined in (\ref{Tdot}),  $ \TT_j
= \mathcal{D}_j \setminus \mathcal{D'}_j$, and $\mathcal{D}_j,
\mathcal{D'}_j$ are defined in Proposition (\ref{l41}),
 and the set
\[
\ddot \TT^2= \bigcup_{
j=-N''_q}^{N'_q} \{ \kk=(k_1, k_2) \in
\dot\TT^2\;, k_2 \in \TT \setminus \mathcal{D'}_j\}.
\]
Hence  the set of   degenerate energies is
$$ \mathcal{ E}'_c = \{ E \in \R, \exists \kk=(k_1,k_2) \in { \TT}^2,
  \exists j = -N''_{q},...,
N'_q, \; E= E_2(\kk)= E_j(k_2) \},
$$
By  Proposition (\ref{l41})  $\mathcal{ E}'_c$ is a discrete set as well as
the set
\[
D'= D \cup  \mathcal{ E}'_c,
\]
where $D$ is defined in
(\ref{D}).

%%%%%%%%%%%%%%%%%%%%%%%%%%%%%%%%%%%%th45%%%%%%%%%%%%%%%%%%%%%%%%%%%%

\begin{theorem}\label{t45}
 Let  $H=H_0 + V$  be  the  operator  defined
 in Theorem \ref{t41}, $G(\x,\y; z)$ be its Green function, and $G(\x,\kk;z)$ be
 defined in (\ref{Gk}).  Then:
\begin{itemize}
\item[(i)]  for $z=E_{2}( \kk)  \mp i\varepsilon $ the limits
\begin{equation}\label {limb}
\Psi_{v,\pm}(\x,\kk) = \lim_{\varepsilon \to +0}
\Psi_{z}(\x,\kk)\Big|_{z=E_2(\kk)
 \mp i\varepsilon}
=\lim_{\varepsilon \to +0} \pm i\varepsilon G(\x,\kk;(E_2(\kk)
\mp i\varepsilon)),
\end{equation}
exist for all  $ \kk \in \ddot \TT^{2}$, are bounded in $ \x \in
\ZZ^2$ for any $ \kk \in \ddot \TT^{2}$, are continuous in $\kk$
on any compact subset of $ \ddot \TT^{2}$ for any $\x \in \ZZ^2$,
and satisfy the Schr\"odinger equation:
\begin{equation}\label{SEKb}
((H_0 + V) \Psi_{v,\pm})(\x,\kk)=E_2(\kk) \Psi_{v,\pm}(\x,\kk);
\end{equation}
\item[(ii)] For $z= E_{j}( k_2) \mp i\varepsilon, \;  j=-N''_{q},...,
N'_q $ the  limits
\begin{eqnarray}\label {lims}
\Psi_{s,j,\pm}(\x,k_2)  =
\lim_{\varepsilon \to +0}
 \mp i\varepsilon
I_j(\kk;z)G(\x,\kk;(E_j(k_2) \mp i\varepsilon)), \nonumber
\end{eqnarray}
in which
\[I_j(\kk;z)=(E_2(\kk)-z)\left[\int _{\TT}dk_1\frac{1}
{\vert(E_2(\kk)- z)\vert^2}  \right]^{1/2}
\]
exist for any  $\kk=(k_1,k_2) \in \TT_j^2$, are bounded in $
\x \in\ZZ^2$ for any
  $k_2 \in\TT_j$,  are continuous in $k_2$ on any compact
subset of $ { \TT}_j$ and
 satisfy the Schr\"odinger equation:
\begin{equation}\label{SEKs}
((H_0 + V) \Psi_{s,j,\pm})(\x,k_2)=E_j(k_2) \Psi_{s,j,\pm}(\x,k_2).
\end{equation}
\item[(iii)] $\Psi_{s,j,\pm}(\cdot,k_2), \; k_2 \in \TT_j$ are  surface states
in the  sense of  Definition 3.1.
\end{itemize}
\end{theorem}
%%%%%%%%%%%%%%%%%%%%%%%%%%%%fin du t45%%%%%%%%%%%%%%%%%%%
\noindent {\it Remarks}. 1). It can be shown that for all  $ \kk \in
\ddot{ \TT}^2$ such that $ E_2(\kk) \in(-d,d) \setminus
\bigcup_{j= -N''_q}^{ N'_q} \mathrm{Ran}E_j $ the function
 $\Psi_{v,\pm}$, defined by (\ref{limb}), is  the unique
 solution of the  integral equation:
\begin{equation}\label{LSb}
 \Psi(\x,\kk)=e^{2i\pi\kk.\x} - \sum_{\y \in \ZZ^d}G_0^{(2)}(\x-\y; E_2
(\kk) \mp i0)V(\y)\Psi(\y,\kk),
\end{equation}
that has to be understood in the same  way as in Theorem
\ref{t33}\textit(ii).
 On the other hand,  it is easy to check that for any $ j=-N''_{q},...,
 N'_q,\; \kk=(k_1,k_2) \in \TT^2_j$
 and $E_j(k_2) \not \in [-d,d]$,
$\Psi_{s,j,\pm}(\x,k_2)$ is a   solution of the homogeneous
integral equation:
\begin{equation}\label{LSs}
 \Psi(\x,k_2)= - \sum_{\y \in\ZZ^d}G_0^{(2)}(\x-\y; E_j(k_2)
 \mp i0)V(\y)\Psi(\y,k_2).
\end{equation}

\smallskip \noindent 2). One can view the above results from the point
of view of the direct integral decomposition technique for  finite
difference operators with periodic coefficients \cite{DS}.Namely by using
the periodicity in $x_2$ of the operator $H$ with $\alpha=p/q$, we
can write the direct integral decomposition
\begin{equation}\label{dir}
H=\int_{|k_2-1/2|\le 1/2q}^{\oplus} H_q(k_2)dk_2.
\end{equation}
Here $H_q(k_2)$ is the selfadjoint operator defined by the
restriction of $H$ to the linear manifold of functions
$\Psi_{k_2}(\x)=e^{2i\pi k_2 x_2}\Phi_{k_2}(\x)$, where
$\Phi_{k_2}$ is $q$-periodic in $x_2$. Thus $H_q(k_2)$ acts in the
strip $\{x_1 \in \ZZ, \; 1 \le x_2 \le q \}$, and is the
perturbation of the respective Laplacian by the $q$ rank
potential (\ref{Vt}) with $1 \le x_2 \le q$. This implies that
the spectrum of $H_q(k_2)$ consists
of two parts. The first is the absolutely continuous component:
the union of values of the functions $-\cos 2\pi k_1 - \cos
2\pi(k_2+l/q), \; k_1 \in \TT, \; l=1,...,q$ and $k_2 \in [1/2 -
1/2q,1/2+1/2q)$ is fixed, the corresponding eigenfunctions are deformed
plane waves in $x_1$. The second part is discrete spectrum,
consisting of $N_q \le q$ eigenvalues $E_j(k_2)$, lying outside of
the above absolutely continuous spectrum, and having exponentially
decaying in $x_1$ eigenfunctions.As $k_2$ varies in the direct integral
the absolutely continuous
spectrum of $H_q(k_2)$ gives rise to the volume states of the
operator $H$, while the discrete spectrum of $H_q(k_2)$ gives rise
to the surface states.

%%%%%%%%%%%%%%%%%%%%%%%%%%%%proof of th45%%%%%%%%%%%%%%%%%%%%%%%%%%%%%%%%%%%%%%%%%%

\medskip \noindent { \it Proof of Theorem \ref{t45}.} Take $ (E, \kk) \in
\sigma(H) \times {\ddot \TT}^2$, $z= E \pm i
\varepsilon, \varepsilon > 0$ and denote $\Psi_{z}(\x,\kk)= (E
-z)G(\x,\kk;z)$, where $G(\x,\kk;z)$ is defined in  (\ref{Gk}). We
know from the proof of Theorem \ref{t33} that  if for any $\x
\in\ZZ^2$ the limit
 $\Psi_{E}(\x,\kk)=\lim_{\varepsilon \to 0} \Psi_{z}(\x,\kk)$ exists,
then
 $\Psi_{E}$
is a solution of the Schr\" odinger equation  $H\Psi_{E}=
E\Psi_{E}$. By Theorem \ref{t41} we can write the representation:
\begin{eqnarray}\label{psiZp}
G(\x,\kk,z)&=&\frac{1}{E_2(\kk)- z}\Big[e^{2i\pi\kk \cdot \x} \\&+&
\sum_{m=0}^q t_m(k_2-m \alpha;z) G_0^{(1)}(x_1;z+ \cos 2\pi
(k_2-m\alpha))e^{2i\pi(k_2-m\alpha)x_2}\Big]. \nonumber
\end{eqnarray}
Choose first a pair $(\kk \in {\ddot \TT}^2, E=E_2(\kk))$, as it
was done in the proof of Theorem \ref{t33} for the quasiperiodic
case. By Proposition \ref{l41}  the denominator
$1-P_q(E,k_2)$ in $t_m$ of (\ref{tmp}) is nonzero and we obtain
from (\ref{psiZp}):
\begin{eqnarray}\label{Sopv}
\Psi_{v,\pm}(\x,\kk)&=&e^{2i\pi\kk.\x} \\&+&\sum_{m=0}^q
t_m(k_2-m \alpha;z) G_0^{(1)}(x_1;z+ \cos2\pi(k_2-m\alpha))\Big|_{z=
E_2(\kk) \mp i0} e^{2i\pi(k_2-m\alpha)x_2}\nonumber
\end{eqnarray}
This proves the first assertion of the theorem.

\medskip \noindent
Consider now the case where $\kk= (k_1, k_2) \in {\TT}^2_j$, and
$E=E_j(k_2)$ for some $j=-N''_{q},..., N'_q$.  We know that the
pair $(\kk,E)$ is such that $E+ \cos 2 \pi (k_2 + l\alpha) \not\in
(-1 ,1)$ for any $l= 1,...,q$.  Hence, by using the separability
property (\ref{sb}) and the periodicity of the $E_j$'s,  given by
Proposition \ref{l41}, we find that  $E+\cos2 \pi (k_2 + l\alpha)
\not\in [-1 ,1]$ for any $l= 1,...,q$, i.e. all that these
energies belong to the resolvent set of the $1$-dimensional
Laplacian.

\medskip \noindent This observation implies the existence of the limit
\[
\Psi_{s,j,\pm}(\x,k_2) = \lim_{\varepsilon \to \mp 0}\big[I(\kk,z)
\Psi_{z}(\x,\kk)\big] \Big|_{z= E_j(k_2) \mp i\varepsilon},
\]
provided that the limit
\begin{equation}\label{lims1}
\lim_{\varepsilon \to +0} \frac{ \varepsilon}{1- P_q(k_2,
E_j(k_2)\mp i\varepsilon )}.
\end{equation}
exists. This can be proved by using the relations
\[
1 - P_q(k_2, E_j(k_2)\mp i\varepsilon)=\pm i \varepsilon
\partial_E P_q(k_2,E_j(k_2)) + O(\varepsilon^2),
\]
valid for sufficiently small $\varepsilon $,  and the relation
$\partial_E P_q \not= 0$. Now, it easy to verify that
\begin{equation}\label{Sops}
 \Psi_{s,j,\pm}(\x,k_2)=I_j(k_2)
\sum_{m=1}^q \tilde t_m(k_2-m \alpha) G_0^{(1)}(x_1;E_j(k_2) +
\cos2\pi(k_2-m\alpha)) e^{2i\pi (k_2-m\alpha)x_2},
\end{equation}
where
\begin{equation} \label{I}
\left.I_j(k_2)=\left[\int _{\TT}dk_1\frac{1}{\vert(E_2(\kk)-
z)\vert^2}  \right]^{1/2} \right|_{z=E_j(k_2) \mp i0},
\end{equation}
$$ \left.\tilde t_1(k_2)={2i \sigma}\Big((g \hat
\gamma_0(k_2;z)+i)
\partial_E P_q(k_2,z) (g \hat
\gamma_0(k_2+\alpha;z)+i))\Big)^{-1}\right|_{z=E_j(k_2) \mp i0},
$$ and  for $ m\geq 2$
\begin{eqnarray} \label{t>2}
\tilde t_m(k_2; E)&=&{2ig \sigma}\Big((g \hat
\gamma_0(k_2;z)+i) \partial_E P_q(k_2,z)(g \hat
\gamma_0(k_2+m\alpha;z)+i)\Big)^{-1}
\\ &\times& P_{m-1} (k_2;z)\Big|_{z=
E_j(k_2) \mp i0}. \nonumber
\end{eqnarray}
By using the same argument as that in the proof of (\ref{SEKb}),
we find that $\Psi_{s,j,\pm}$ satisfies (\ref{SEKs}). Let us prove
now that $\Psi_{s,j,\pm}$ is a surface state. We  know that
$E+\cos 2 \pi (k_2 +l\alpha) \not\in [-1 ,1]$ for any $l=
1,\cdots,q$.
 Since   all these energies  are in the resolvent set of the
 $1$-dimensional Laplacian, each term of  the sum of the r.h.s. of
(\ref{Sops}) decays exponentially with respect the transverse
coordinate $x_1 \in\ZZ$. Since the number of these terms is
finite, we conclude that for any $x_2 \in \ZZ$,
$\Psi_{s,j,\pm}(.,x_2) \in l^2(\ZZ)$. The proof of the theorem is
complete.
%%%%%%%%%%%%%%%%%%%%%%%%%%%%%%%%%%%%%%%%%%%%%%%%%%%%%%%%%%%%%%%%%%%

\medskip \noindent We can now use the last theorem, where we have
constructed the generalized eigenfunctions (\ref{Sopv}) and
(\ref{Sops}), to prove Theorem \ref{t42}.

\medskip \noindent {\it Proof of Theorem }\ref{t42}.  It follows
from the proof of the Theorem  \ref{t43}  that
 $\sigma (H) \subset  [-d,d] \cup ( \cup_{j=-N"_q}^{
N'_q} { \rm RanE_j}) $. Hence we have to prove the opposite
inclusion. For the part $[-d,d]$ of the spectrum the inclusion was
proved in Theorem \ref{t22}.  So assume that  $E \in
\bigcup_{j=-N"_q}^{ N'_q}{ \rm Ran }E_j \setminus [-d,d]$ is such that
there exists  a surface state $\Psi_{s}(\x)$
 satisfying the Schr\"odinger equation: $
(H \Psi_{s})(\x)=E \Psi_{s}(\x)$.
 We  apply again the H.~Weyl criterion, setting
$$\Psi_n(\x)=1_n(x_2)\Psi_{s}(\x)/ N_n ; \quad N_n= \Vert 1_n\Psi_{s}
\Vert _{l^2(\ZZ^2)}, $$ where $ 1_r$  is the indicator of the ball
$\{ x_2 \in \ZZ: \;  \vert x_2 \vert \leq n \}$. A
straightforward calculation shows that $ C_1 n^{1/2} \leq  N_n \leq C_2
n^{1/2} \quad \hbox{as}\quad n \to \infty$ for some strictly positive
constants $C_{1,2}$, and that
 $$ (H\Psi_n)(\x)=\left\{
\begin{array}{l} E\Psi_n(\x), \quad \vert x_2 \vert < n;\\
A_n(\x), \quad n \leq \vert x_2 \vert \leq n+1;\\ E\Psi_n(\x)= 0,
\quad \vert x_2\vert \geq n +2,
\end{array}
\right. $$ 
where $\Vert A_n \Vert _{l^2(\ZZ^2)} = O(n^{-1/2}), \quad
n \to \infty$. It is  easy to check that $\Psi_n$ is a Weyl
sequence for $H$ at the energy $E$. This proves the theorem.

\medskip \noindent Our next result  concerns the completeness of
the system of generalized eigenvectors  (\ref{Sopv}) and
(\ref{Sops}), defined in Theorem  \ref{t45}.

%%%%%%%%%%%%%%%%%%%%%%%%%%%%%%%%%%%%theorem46%%%%%%%
\begin{theorem}\label{t46}
 Let  $H=H_0+V$  be the selfadjoint operator in $l^2(\ZZ^2)$ defined
 in Theorem \ref{t41}.
 Consider  the family $ {\cal F}=\{ \Psi_{v}(\x,\kk), \; \x \in \ZZ^2\}_{\kk
 \in {\ddot \TT}^2}   \bigcup_{j=-N''_q}^{N'_q}\{ \Psi_{s,j}(\x,k_2);
\x \in
\ZZ^2\}_{k_2 \in  \TT_j} $, defined by (\ref{Sopv}) and
by (\ref{Sops}). Then ${\cal F}$ is a complete system of
generalized eigenfunctions of $H$ in  any  sufficiently small
interval $\Delta$ of $ \sigma(H)$ such that $\Delta \cap D' =
\emptyset$, i.e.
\begin{itemize}
\item[( i)]  for any $f \in l^2(\ZZ^2)$  the series
\[F_v(\kk)=\sum_{\x \in \ZZ^d} \overline {\Psi_v(\x,\kk)}f(\x), \; \kk \in
\ddot \TT^2,
\]
and
\[\quad  F_{s,j}(k_2)=\sum_{\x \in \ZZ^d} \overline {
\Psi_{s,j}(\x,k_2)}f(\x), \; k_2 \in  \TT_j, \; j=
-N''_{q},...,N'_q\] converge in $l^2(\ZZ^2)$;
\item[(ii)] if ${\cal E}_H (\Delta)$ is the spectral projection of $H$
 corresponding to the  interval $\Delta \in \sigma(H)$,
 then $$ \Vert {\cal E}_H
(\Delta)f\Vert^2=\int_{ \{\kk \in \ddot\TT^2: E_2(\kk)
 \in \Delta \}} \vert
 F_v(\kk) \vert^2d \kk + \quad \sum_{j=-N''_q}^{N'_q}  \int_{ \{ k_2 \in
 \TT_j: E_j(k_2) \in \Delta \}}
 \vert F_{j,s}(k_2) \vert^2d k_2;$$
\item[(iii)] for the same interval we have
\begin{eqnarray*} \Vert H{\cal E}_H  (\Delta)f\Vert^2&=&\int_{ \{ \kk \in \ddot
 \TT^2: E_2(\kk) \in \Delta \} } \vert E_2(\kk)\vert^2 \vert F_v(\kk)
\vert^2d \kk \\&+& \quad \sum_{j=-N''_q}^{N'_q} \int_{ \{ k_2 \in
 \TT_j: E_j(k_2) \in \Delta \}}
 \vert E_j(k_2) \vert^2
 \vert F_{s,j}(k_2) \vert^2 dk_2.
 \end{eqnarray*}
\end{itemize}
\end{theorem}
%%%%%%%%%%%%%%%%%%%%%%%%%%%%%%%%%%%%%%%%%%%%%%%%%%%%%%%%%%%%%%%%%%%%
\noindent{\it Proof.} For any  compact interval $\Delta
\subset
\sigma(H) \setminus D'$ consider  the sets:
\begin{equation}\label{nunu}
\nu = \{ \kk \in \ddot \TT^2: E_2(\kk) \in \Delta\}, \quad
\nu_j=\{ \kk = (k_1,k_2)\in \TT^2_j: E_j(k_2) \in \Delta\},
j=-N''_q,...,N'_q.
\end{equation}
Proposition \ref{l41} implies that there exists a constant $\eta
>0$ such that: $$  \min_j \inf _{\kk \in \nu \cap \TT_j^2}
\vert E_2(\kk)- E_j(k_2)\vert, \;
 \min_j \inf _{\kk \in \nu_j \cap\ddot \TT^2} \vert E_2(\kk)-
E_j(k_2)\vert \geq \eta.$$ Notice that $\eta $ depends only on the
${\rm dist}( \Delta, D')$. Moreover, if $\Delta $ is sufficiently
small,  then the sets $\nu, \nu_j, j=-N''_q,..., N'_q$ are
disjoint. The subsequent argument uses this property of $\nu$, and
$\nu_j, j=-N''_q,..., N'_q$.

\medskip \noindent
We will follow now the proof of Theorem \ref{t34}. Hence we have
to prove assertion $(ii)$ first for a function
  $f$  with  compact support.  We have for  $z=E+i\varepsilon$,
where  $ E \in \Delta$ and $\varepsilon>0$:
\begin{eqnarray}\label{ImG}
\frac{1}{\pi}\Im(G(z)f,f)&=& \frac{1}{\pi} \int_{\nu} d \kk
\frac{\varepsilon}{\vert(E_2(\kk)-z)\vert^2} \vert F_{z}(\kk)
\vert^2 \nonumber \\ &+& \frac{1}{\pi} \sum_{j=-N''_q}^{N'_q}
\int_{ \nu_j} d\kk \frac{\varepsilon}{\vert(E_j(k_2)-z)\vert^2}
\vert F_{z}(\kk) \vert^2 + O( \varepsilon),
 \end{eqnarray}
where
\begin{equation}\label{hZ}
F_z(\kk)=\sum_{\x \in \ZZ^d} \overline{\Psi_z(\x,\kk)}f(\x), \quad
\Psi_z(\x,\kk)=(\tilde E-z) G(\x,\kk;z),
\end{equation}
 $\tilde E = E_2(\kk)$ or $\tilde E = E_j(k_2)$ and $G(\x,\kk;z)$ is
 defined by (\ref{Gk}) and (\ref{GFp})-(\ref{tmp}).
 Since   for every  $ \kk=(k_1,k_2) \in \nu$  and $ E \in  \Delta  $,
$ P_q(k_2, E) -  1 $   is not zero,  the limit
 $\lim_{\varepsilon \to 0}\Psi_{E+i\varepsilon}(\x,\kk)$ exists
 for any $\x \in \ZZ^2$
uniformly in $ \kk=(k_1,k_2) \in \nu$ and in $ E \in \Delta $.
 Applying to the first term of the r.h.s of (\ref{ImG})
 the operation
$\lim_{\varepsilon \to 0} \int_{\Delta}...dE$, we get:
 \begin{equation}\label{lim1}
 \lim_{\varepsilon \to 0}\int_{\Delta}  dE
\frac{1}{\pi} \int_{\nu} d \kk \frac{\varepsilon}
{\vert(E_2(\kk)-z)\vert^2}  \vert F_{z}(\kk) \vert^2=\int_{ \{\kk
\in \ddot\TT^2: E_2(\kk) \in \Delta \}}
 \vert F_v(\kk) \vert^2d \kk.
\end{equation}
 So we are left with the second term of the r.h.s of (\ref{ImG}).
 For every  $j \in \{-N''_q,...,N'_q\}$, $\kk=(k_1,k_2) \in \nu_j$, and
$ E \in \Delta  $,
 we have:
\begin{equation}\label{426}
\lim_{\varepsilon \to 0} F_{E+i\varepsilon}(\kk) = \sum_{\x \in
\ZZ^d}  \overline{\Psi(\x,\kk;E)}f(\x),
\end{equation}
where
\begin{eqnarray} \label{Psi}
\Psi(\x,\kk,E)&=&\frac{(E_j(k_2)-E)}{E_2(\kk)- E}
\Big[e^{2i\pi\kk\cdot \x} \\ &+& \sum_{m=0}^q t_m(k_2-m \alpha;E+i0)
G_0^{(1)}(x_1;E+i0 + \cos2\pi(k_2-m\alpha))e^{2i\pi
(k_2-m\alpha)x_2}\Big], \nonumber
\end{eqnarray}
 which  in
particular corresponds to $[(E_2(\kk)- E_j(k_2)) I_j(k_2)]^{-1}
\Psi_{s,j}(\x, k_2)$ for $E=E_j(k_2)$. The limit (\ref{426}) is
also
 uniform in $ \kk=(k_1,k_2) \in \nu_j$  and in $ E \in \Delta $.
Applying again the same operation:
$
\lim_{\varepsilon \to 0} \int_{\Delta} ... dE$ to  the $j$th term
of the sum in r.h.s of \ref{ImG}, we get
\begin{equation}\label{lim2}
 \lim_{\varepsilon \to 0}\int_{\Delta}  dE
\frac{1}{\pi} \int_{\nu_j} d \kk \frac{\varepsilon}
{\vert(E_j(k_2)-z)\vert^2}  \vert F_{z}(\kk) \vert^2 = \int_{ \{
k_2 \in  \TT_j: E_j(k_2) \in \Delta \}}
 \vert
 F_{j,s}(k_2) \vert^2d k_2.
\end{equation}
Relations (\ref{lim1}), and (\ref{lim2}) imply assertions
$(i)$ and $(ii) $ of the theorem for the case of a function $f$ with
compact support. The proofs of these assertions for an arbitrary
function $f \in l^2(\ZZ^2)$, and the proof of assertion $ii)$
require standard means of spectral theory (see the proof of
Theorem \ref{t34}).

%%%%%%%%%%%%%%%%%%%%%%%%%%%%%preuve du Prop41%%%%%%%%%%%%%%%%%%%%%%%%%%%%%%%%%
\medskip \noindent
\textit{Proof of Proposition} \ref{l41}
 According to (\ref{Pm}),
 we can write equation (\ref{be}) for $\alpha=p/q$ as
\begin{equation}\label{be1}
P_q(k_2,E)=\sigma ^{q}\prod_{l=1}^{q}\widehat{b}(k_{2}+l/q,E+i0)=1,
\end{equation}
where
\begin{equation} \label{bkG}
\widehat{b}(k_{2},z)=\frac{gG_0^{(1)}(0,z+ \cos 2\pi k_{2})-i}
{gG_0^{(1)}(0,z+\cos 2\pi k_{2})+i}%.
\end{equation}

 \medskip \noindent Since the product in the l.h.s. of the equation \ref{be1}
 is periodic in
$k_{2}$ with period $1/q$, its solutions are also periodic in
$k_{2}$ with period $1/q$, and we can restrict ourselves to the
interval $[1/2-1/(2q),1/2+1/(2q))$. By Lemmas \ref{lA1} -
\ref{lA2} we have
 $|\widehat{b}(k_{2},z)|\leq 1,\;k_{2}\in \TT,\;z\in
\mathbb{C}$, thus the band equation (\ref{be1})
admits a solution if and only if the modulus of each factor $\widehat{b}%
(k_{2}+l/q,E+i0),\;l=1,...,q$ in its l.h.s is 1. Hence, we can
write the representation
\[
\widehat b(k_{2},E)=\exp \{2\pi i\phi (k_{2},E)\}.
\]
In what follows we will consider the case where $E$ is positive
(the arguments for negative $E$ are similar and will be omitted).
In this case we have from (\ref{G01}):
\[
G_{0}^{(1)}(0,E+i0)=-\frac{1}{\sqrt{E^{2}-1}},\;E>1,
\]
and we can choose the phase $\phi (k_{2},E)$ as
\begin{equation}\label{phi}
\phi (k_{2},E)=1/\pi \arctan \big((1/g)\sqrt{(E+\cos 2\pi
k_{2})^{2}-1}\big).
\end{equation}
\medskip \noindent For any $k_{2}\in [1/2-1/(2q),1/2+1/(2q)) \quad \phi$ is
 a non-negative and an increasing function of
$E\geq 1-\cos 2\pi k_{2} $,
satisfying the inequalities:
\[
0\leq \phi (k_{2},E)<1/2.
\]
The above formulas show that the l.h.s of equation (\ref{be1}) is
real analytic in the domain
\[
\{(k_{2},E):k_{2}\in [1/2-1/(2q),1/2+1/(2q)),\; E>1-\cos 2\pi
k_{2}\},
\]
hence solutions of the equation, if they exist,  are real analytic
in $k_{2}$ (notice that  here the condition $E>1-\cos 2\pi k_{2}$
is equivalent to the conditions $E>1-\cos 2\pi( k_{2}+ l\alpha),
\forall l =1,..,q$).

\medskip \noindent We will use (\ref{be1})
in the form
\begin{equation}
\Phi _{q}(k_{2},E)-q\omega =0\;(\mathrm{mod\;}1),  \label{Pha1}
\end{equation}
where
\begin{equation}
\Phi _{q}(k_{2},E)=\sum_{l=1}^{q}\phi (k_{2}+l\alpha,
E)=\sum_{l=0}^{q-1}\phi (k_{2}+l\alpha ,E). \label{Pha0}
\end{equation}
For any fixed $k_{2}\in [ 1/2-1/(2q),1/2+1/(2q))$,  $\Phi
_{q}(k_{2},E)$ is a positive and an increasing function of $E \geq
1-\cos 2\pi k_{2}$, bounded by $q/2$.

\medskip \noindent Fix now $q$ and $\omega$  and  denote
by $\alpha _{\omega }$
the integer part of the minimum
\[
\min_{ k_{2}\in [1/2-1/(2q),1/2+1/(2q))}
\Phi_{q}(k_{2},1-\cos 2\pi k_{2})-q\omega.
\]
For a fixed integer $j$ denote by   $E_{j}(k_{2})$  the energy
such that
\begin{equation}\label{Ej}
 \Phi _{q}(k_{2},E_{j}(k_{2}))-q\omega =\alpha _{\omega }+j
\end{equation}
and  denote by $\mathcal{D}_{j}$ the set of $k_2 \in \TT$ such
that (\ref{Ej})
 is satisfied. The sets $\mathcal{D}_{j}, j=-N''_q,...,N'_q$ form
 an increasing family of the closed subset of $\TT$. For all $j$
larger than some $j_0$ , $\mathcal{D}_{j}$ coincides
with  $ \TT$.

\medskip \noindent Hence $E_{j}$ is the $j$-th  energy band
function and  Ran$E_{j}$ is  the $j$-th surface energy band. It is
clear that the maximum value $ N_{q}^{\prime }$ of $j$ for which
such a solution exists, is such that $N_{q}^{\prime }\leq q/2$.

\medskip \noindent
Since $\Phi _{q}(k_{2},1-\cos 2\pi k_{2})-q\omega $ is analytic in
$k_{2}\in \TT$, it may exist a discrete set $\mathcal{D}'_{j}$ of $
k_{2} \in \TT$, for which $\Phi
_{q}(k_{2},1-\cos 2\pi k_{2})-q\omega) $ is equal to the integer
$\alpha _{\omega }+j$.
Numerical experiments show that for small $q$ there are at most
two values of $k_{2}$ in the interval
$[1/2-1/{(2q)},1/2+1/{(2q)})$ for which this
event occurs, so the number of points in  $\mathcal{D}'_{1}$
is $2q $ and the other
$\mathcal{D}'_{j}$ are empty.

\medskip \noindent
We have proved that if $k_{2}\in \TT_j=\mathcal{D}_{j}\setminus
\mathcal{D'}_{j}$,   $E_{j}(k_{2})$ exceeds $1-\cos 2\pi k_{2}$,
then we have for all $\mathbf{k}\in \TT^2_j$:
\begin{equation}
E_{j}(k_{2})>1-\cos 2\pi k_{2}\geq -\cos 2\pi k_{1}-\cos 2\pi k_{2}=E_{2}(
\kk),  \label{in3}
\end{equation}
i.e. the separation property (\ref{sv}) between the band of the
volume states and the surface bands.

\medskip \noindent
Let us now discuss separation between the surface bands $
E_{-N''_q},...E_{N_{q}^{\prime }}$. We will use the relation
\begin{equation}
1=\Phi _{q}(k_{2},E_{j+1}(k_{2}))-\Phi _{q}(k_{2},E_{j}(k_{2})),
\label{dif}
\end{equation}
implied by (\ref{Pha1}).

\medskip \noindent
Consider first   the energy range  $E\geq \epsilon_{m}$ for some $\epsilon_{m}>2$.
It follows from
(\ref{Pha1}) that the maximum energy $E_{q}$ for which the
equation is soluble is finite (this is the upper
edge of the spectrum of the operator $H$ for a given $q$). Hence
the partial derivative
\begin{equation}
 \frac{{\ \partial \Phi _{q}}}{{\partial E}}=1/\pi
 \sum_{l=0}^{q-1}\frac{{g}}{%
{g^{2}+[E+\cos 2\pi (k_{2}+l/q )]^{2}-1}}\frac{{E+\cos 2\pi (k_{2}+l/q)}%
}{\sqrt{(E+\cos 2\pi (k_{2}+l/q ))^{2}-1}}  \label{pder}
\end{equation}
 satisfies the inequalities:
\[ 0 <
\frac{{\ \partial \Phi _{q}}}{{\partial E}}\leq \frac{q}{\pi g}\frac{E_{q}{+1%
}}{\sqrt{(\epsilon_{m}-2)\epsilon_{m}}}:=(\eta _{q}^{\prime })^{-1}<\infty .
\]
This bound and (\ref{dif}) lead to the relations
\begin{equation}
1=\int_{E(k_{2})}^{E_{j+1}(k_{2})}{\frac{{\ \partial \Phi _{q}}(k_{2},E)}{{%
\partial E}}}dE\leq (E_{j+1}(k_{2})-E_{j}(k_{2}))(\eta _{q}^{\prime })^{-1},
\label{ineq}
\end{equation}
implying the separation property (\ref{sb}) in the case where
$E_{j}(k_{2})>2 $.

\medskip \noindent
In the case, where
\[0\leq 1-\cos 2\pi k_{2}\leq E_{j}(k_{2})\leq
2,\;k_{2}\in \lbrack 1/2-1/(2q),1/2+1/(2q)),
\]
the r.h.s of (\ref{pder}) can be infinite because of the
contribution of the first term (for $E=1-\cos 2\pi k_{2}$), and of
the second term (for $E=1-\cos 2\pi k_{2}$, and
$k_{2}=1/2-1/(2q)$) or of the $(q-1) $th term (for $E=1-\cos 2\pi
k_{2}$, and $k_{2}=1/2+1/(2q)$). Since, however,  each term in the
phase (\ref{Pha0}) is non-negative and
\[
\phi_{q}:=\max_{ k_{2}\in [1/2-1/(2q),1/2+1/(2q)) } \phi (k_{2},2)
\]
 is strictly less than $1/2$, the
contribution of these terms in the difference (\ref{dif}) is
bounded from above by $2\phi _{q}<1$, and we obtain from
(\ref{dif}) the inequality
\[
0<1-2\phi _{q}<\int_{E_{j}(k_{2})}^{E_{j+1}(k_{2})}{\frac{{\ \partial }%
\widetilde{\Phi }_{q}(k_{2},E)}{{\partial E}}}dE,
\]
where $\widetilde{\Phi }_{q}(k_{2},E)$ is the sum in (\ref{Pha0}),
in which the terms corresponding to $l=0$ and to $l=1$ if
$k_{2}\in \lbrack 1/2-1/(2q),1/2)$, and to $l=q-1$ if $k_{2}\in
\lbrack 1/2,1/2+1/(2q))$ are omitted. It is easy to check that the
partial derivative of $\widetilde{\Phi }_{q}(k_{2},E)$ with
respect to $E$ is bounded from above by a constant $(\eta
_{q}^{\prime \prime })^{-1}<\infty $. This leads to the bound
(\ref{ineq}) in which $(\eta _{q}^{\prime })^{-1}$ is replaced by
$(\eta _{q}^{\prime \prime })^{-1}$ and ${\Phi }_{q}$ by
$\widetilde{\Phi }_{q}$. The obtained bounds imply the separation
property (\ref{sb}) with $\eta _{q}=\min \{\eta _{q}^{\prime
},\eta _{q}^{\prime \prime }\}$. Proposition \ref{l41} is proved.
%%%%%%%%%%%%%%%%%%%%%%%%%%%%%%%%%%%%%%%%%%%%%%%%%%%%%%%%%%%%%%%%%%%%%%%%%%%%%%

\medskip \noindent \textit{Remark}.
It can be seen from the proof above that the
distance between the bands  increases as $|j|$
increases. Besides, the  distance between the two first
bands  is   of order  $O(1/q)$ when $q$ is large.

\medskip  \noindent
Denote from now  on the operator of (\ref{ham}) - (\ref{vt}) as
$H_{\alpha }$.
 We conclude this section by discussing
correspondence between the spectrums of the operators
$H_{\alpha }$ with an irrational number $ \alpha $
and  with its rational approximations
$p_{n}/q_{n}$:
\begin{equation}
\lim_{n\rightarrow \infty }\frac{p_{n}}{q_{n}}=\alpha .
\label{pqa}
\end{equation}
It is easy to prove, by using the basic formula (\ref{Gs}) for the
resolvent, that
under condition (\ref{pqa}) $H_{p_{n}/q_n}$ converges to $%
H_{\alpha }$ in the strong resolvent sense. Hence, according to
general principles \cite{Ka}, the spectrum $\sigma
(H_{p_{n}/q_{n}})$ is upper semi-continuous in $n$ in the limit
(\ref{pqa}). Here is a statement, that gives a more detailed
behavior of $\sigma (H_{p/q})$ for large $q$. Recall, that $\sigma
(H_{p/q})$ is the union of the interval $[-2,2]$ and of
$N''_{q}+N'_{q}$ surface bands, part of which can belong or
intersect the interval $[-2,2]$.

%%%%%%%%%%%%%%%%%%%%%%%%%%%%%%t52%%%%%%%%%%%%%%%%%%%%%%%%%%%%%%%%%%%%%%%%

\begin{theorem}\label{52}
Assume that q sufficiently large. Then there exists at most one
negative surface  energy band above $E=-2$ and at most one
positive surface energy  band below $E=2$. These bands, if they
exist, have the width of  order $O(1/q^{2})$ as $q \to \infty$.
The width of the surface  energy bands lying in $(-\infty ,-2) $
and in  $(2,+\infty )$, are of order $O(\exp \{-\emph{const}\cdot
q\})$ as $ q\rightarrow \infty $.
\end{theorem}
%%%%%%%%%%%%%%%%%%%%%%%%%%%%%%%%%FIN du t52%%%%%%%%%%%%%%%%%%%%%%%%%%%%%%
\noindent \textit{Proof}.
We start from the dispersion equation (\ref{Pha0}%
) - (\ref{Pha1}) for the surface  energy bands. Since  the function $\Phi
_{q}(k_{2},E)$ has period $1/q$ in $k_{2}$, its Fourier series is:
\[
\Phi _{q}(k_{2},E)=\sum_{n\in \ZZ}\widehat{\Phi
}_{q,n}(E)e^{-2\pi ik_{2}qn},
\]
where
\begin{eqnarray*}
\widehat{\Phi }_{q,n}(E)= q\int_{0}^{1/q}dk_{2}e^{2\pi
ik_{2}qn}\phi (k_{2}+l/q,E) = q\int_{0}^{1}dk_{2}e^{2\pi
ik_{2}qn}\phi (E,k_{2}):=q \hat{\phi}_{qn}(E),
\end{eqnarray*}
and $\hat{\phi}_{qn}$ is the $qn$-th Fourier coefficient of the
function $\phi(.,E)$. Hence
\begin{equation} \label{phiq}
\Phi_{q}(k_{2},E)=q\Big(\hat{\phi}_{0}(E)+\hat{\phi}_{q}(E)e^{-2\pi
ik_{2}q}+...   \Big).
\end{equation}
According to (\ref{phi}), the function $\phi (\cdot,E)$ is
analytic for $|E|> 2$, thus  its Fourier coefficient $\hat{\phi}
_{q}(E)$ is of order $\exp \{-\emph{const}\cdot q\}$ as
$q\rightarrow \infty $.   In addition, formula (\ref{Pha1})
implies the relation
\[
{\frac{{dE_{j}}}{{dk_{2}}}}={\frac{{\partial \Phi _{q}}}{{\partial k_{2}}}}%
\cdot \left({\frac{{\partial \Phi _{q}}}{{\partial E}}}\right)^{-1}.
\]
\medskip \noindent  It follows now from (\ref{phiq}) and from the exponential
decay of the Fourier coefficient
 $\widehat{\Phi }_{q,n}(E) $ that the upper bound for
the derivative $ \displaystyle \frac{{\partial \Phi
_{q}}}{{\partial k_{2}}}$
is of order $O(q^{2}e^{-\mathrm{const} \cdot q})$, while the lower bound for
$\displaystyle \frac{{%
\partial \Phi _{q}}}{{\partial E}}$, which is reached for the highest
energy band $E_{N_{q}^{\prime }}(k_{2})$,  is   of order   $O(q)$.
Thus the derivative  $\displaystyle \frac{{dE_{j}}}{{dk_{2}}}$ is of order
$O(qe^{-\mathrm{const}\cdot q})$.
Since $E_{j}(k_{2})$ is periodic in $k_{2}$
with period $1/q$, then
denoting respectively by  $E_{j}^{max}$, and by $E_{j}^{min}$ the maximum
and the minimum of the $j$-th band function $E_{j}$, we see that
 $|E_{j}^{max}-E_{j}^{min}|$ is  of the order  $\exp
\{-\mathrm{const}\cdot q\}$ if $ E_{j}^{min}> 2$.

\medskip \noindent
Let us fix $ k_{2}\in \lbrack 1/2-1/(2q),1/2+1/(2q))$. To see how
many bands are in between the lowest possible energy $E=1-\cos 2
\pi k_{2}$ and the energy, $E=2$, let us calculate $\delta \Phi =
\Phi _{q}(k_2,2)-\Phi_{q}(k_2,1-\cos 2 \pi k_2)$. We have:
\[
\delta \Phi  =1/\pi \sum_{l=0}^{q-1}\int_{1-\cos 2\pi k_2}^{2}dE{\frac{g}{{%
g^{2}+(E+ \cos 2\pi ( k_2+l/q))^{2}-1}}}
{\frac{{E+\cos2\pi ( k_2+l/q)}}{\sqrt{(E+\cos2\pi ( k_2%
+l/q))^{2}-1}}.}
\]

\medskip \noindent Performing the integration for the different values of
$l$ and summing respective contributions we obtain that  $\delta
\Phi $ is of the  order   $(1/q)\log q$ as $q\rightarrow \infty $, thus
$\delta \Phi \rightarrow 0$ as $q\rightarrow \infty $. Remembering
that for each $k_{2}$ the energy of a band corresponds to an
entire value of  $\Phi _{q}+q\omega $, we deduce that  for large
$q$ there is at most one band in the interval to the  left of
$2$. Since the minimum of  $E_{1}$ for $ k_{2}\in \lbrack
1/2-1/(2q),1/2+1/(2q))$ is larger than $ 1+\cos \pi /q$, the width of
any  band, lying inside the interval $[1+cos\pi /q,2)$, is bounded
by $\pi^2/2q^{2}$.

\medskip  \noindent
It can also occur that
$E_{1}^{max}>2$. In this case, the same argument as above show that the part
of the energy  band in
$(2, \infty)$  is exponentially small in $q$. Thus the total width
in that case is at most of the order $1/q^{2}$.

\medskip \noindent \textit{Remark}.
The assertion of the theorem can be interpreted  as a kind of
continuity of the spectrum  with respect to the limiting
transition (\ref{pqa}). Indeed, according to the theorem, the
width of the surface bands of $H_{p_n/q_n}$, lying outside the
interval $[-d,d]$, is exponentially small in $q_n \to \infty$. It
can also be shown that the gaps between these bands are of the
order $1/q_n$. This is in agreement with the ``limiting'' form of
this part of the  spectrum of $H_{\alpha}$ for irrational
$\alpha$'s, satisfying the Diophantine condition (\ref{di}).
Indeed, according to \cite{KP}, the spectrum of $H_{\alpha}$ in
this case is pure point and dense on $\mathbb{R} \setminus
[-d,d]$.  Here is one more manifestation of this continuity.

\medskip  \noindent
Recall that according to \cite{KP} the eigenvalues of $H_{\alpha}$ outside
$[-d,d]$ are indexed by $x_2 \in \mathbb{Z}^{d_2}$, and for each
$x_2 \in \mathbb{Z}^{d_2}$ the eigenvalue $E_{x_2}$ is the unique
solution of the equation
\begin{equation}\label{pupos}
f(E_{x_2}) \equiv \alpha \cdot x_2 + \omega \; (\mathrm{mod} \; 1),
\end{equation}
where $f: \mathbb{R} \setminus [-d,d] \to \mathbb{R}$
is the monotone increasing function, defined for $E>d$
as
\begin{equation}
f(E)=-\frac{1}{\pi} \int_{\mathbb{T}^{d_2}}dk_2
\; \mathrm{arctan}\Big(g \gamma_0(k_2,E) \Big)^{-1},
\end{equation}
or, in view of (\ref{Gag}) and (\ref{G01}), and for $d_2=1$
\begin{equation}
f(E)=-\frac{1}{\pi} \int_{\mathbb{T}^{1}}dk_2
\; \mathrm{arctan} \; g^{-1} \sqrt{(E+\cos 2\pi k_2)^2-1}.
\end{equation}
On the other hand, we can write the band equation (\ref{Pha1}) as
\begin{equation}
\frac{1}{q_n}\Phi_q(k_2,E)=\frac{l}{q_n}+\omega
\end{equation}
for some integer $l$. Choosing $l$ in the form
$l=p_n x_2 + m q_n$ for some integer $m$, we can write
the last equation as
\begin{equation}
\frac{1}{q_n}\Phi_{q_n}(k_2,E)=\frac{p_n}{q_n} x_2 +\omega.
\end{equation}
Recalling now the expression  (\ref{Pha0}) for the function
$\Phi_q(k_2,E)$, we conclude that for the limiting transition
(\ref{pqa}) and $E>2$ the equation (\ref{Pha1}), defining the
surface bands of $H_{p_n/q_n}$ outside $[-d,d]$, converges to the
equation (\ref{pupos}), defining the all eigenvalues of
$H_{\alpha}$ for a Diophantine $\alpha$ outside $[-d,d]$.

%%%%%%%%%%%%%%%%%%%%%%%%%%APP%%%%%%%%%%%%%%%%%%%%%%%%%%%%%%%%%%%%%%%%%%%%%%

\section{Auxiliary Facts}

\setcounter{equation}{0}

We present here useful facts on the Green function (\ref{Go}) of the $\nu $%
-dimensional Laplacian and on related quantities.
\begin{lemma}
\label{lA1} Let $G_{0}^{(\nu )}(\x-\y;z),\;x,y\in
\mathbb{Z}^\nu,\Im z\neq 0,$ be
the Green function (\ref{Go}) of the $\nu $-dimensional Laplacian (\ref{lap}%
). Write
\begin{equation}
G_{0 }^{(\nu)}(0;z)=R_{\nu }(z)+iI_{\nu }(z),\;R_{\nu },I_{\nu }\in \mathbb{R%
}.  \label{GRI}
\end{equation}
Then
\begin{itemize}
\item[(i)]  for any $\varepsilon >0$, and $E \in  \mathbb{R}$%
\begin{equation}
|R_{\nu }(E+i\varepsilon )|<\infty ,\;0<I_{\nu }(E+i\varepsilon
)<\infty; \label{RI}
\end{equation}
\item[(ii)]  the limits $R_{\nu }(E+i0)$ and $I_{\nu }(E+i0)$ exist for $%
|E|\neq \nu $, satisfy inequality (\ref{RI}) for $|E|<\nu $, and $%
I_{\nu }(E+i0)=0$ if and only if $|E|>\nu $.
\end{itemize}
\end{lemma}
\noindent \textit{Proof.} The part $(i)$ of the lemma follows from
the integral representation (\ref{G0}). It is also easy to prove
that the limits $R_{\nu }(E+i0)$ and $I_{\nu }(E+i0)$ exist and
are finite for $|E|\neq \nu $ (in fact, for $\nu \geq 3$ they are
finite even for $|E|=\nu $, see Lemma \ref{lA4} below). Thus
we have to prove that $I_{\nu }(E+i0)$ is strictly positive for
$|E|<\nu $. By using (\ref{Go}), it easy to show
that for $\nu =1$
\begin{equation*}
\pi ^{-1}I_{1}(E+i0) := \rho _{1}(E)=\left\{
\begin{array}{c}
(1-E^{2})^{-1/2},\ |E|<1, \\ 0, \ \ \ |E|>1,
\end{array}
\right.
\end{equation*}
and that $\pi ^{-1}I_{\nu }(E+i0)$ is the $\nu $th convolution of $\rho _{1}$.
 These two observations imply
that $\pi ^{-1}I_{\nu }(E+i0)$ is strictly positive if $|E|<\nu $, and is zero for $|E\dot{|}>\nu $.
Lemma is proved.

\begin{lemma}
\label{lA2} Let $\gamma _{0}(z)$ be the operator in
$l(\mathbb{Z}^{d_{2}})$, defined as
\begin{equation*}
\gamma_{0}(z)=P_{\mathbb{Z}^{d_{2}}}G_{0}^{(d)}(z)P_{\mathbb{Z}^{d_{2}}},
\quad d_2 <d,
\end{equation*}
and
\begin{equation}
b(z)=\frac{g\gamma _{0}(z)-i}{g\gamma _{0}(z)+i}.  \label{b1}
\end{equation}
Then the operator $\gamma_0(z)+i$ is invertible for $\Im z \geq 0$,
and the operator $b(z)$ is a contraction for $\Im z\neq 0$:%
\begin{equation*} \label{b<1}
||b(z)||<1.
\end{equation*}
\end{lemma}
\noindent \textit{Proof}. According to (\ref{Go}) and
(\ref{GaxGak}) $\gamma_0(z)$  is the a convolution operator in
$l^2(\ZZ^{d_2})$ and its symbol $\hat \gamma_0(k_2;z)$ satisfies
the inequality: $\Im\hat \gamma_0(k_2;z)\geq 0, \; \Im z \geq 0$.
Since the symbol of $\gamma_0(z)+i$ is $\hat \gamma_0(k_2;z)+i$,
we have that $|\hat \gamma_0(k_2;z)+i|\geq \Im \big(\hat
\gamma_0(k_2;z)+i)\geq 1$. Hence $\gamma_0(z)+i$ is invertible and
$||(\gamma_0(z)+i)^{-1}||\leq 1$.

\medskip \noindent The operator $b(z)$ is a rational function of $\gamma_0(z)$, thus
it norm can be found as
\begin{equation*}
||b(z)||=\sup_{k_{2}\in \mathbb{T}^{d_{2}}}|\widehat{b}(k_{2};z)|.
\end{equation*}
By using (\ref{GRI}), we obtain that
\begin{equation}
\left| \widehat{b}(k_2,z)\right| =\left. \frac{R_{d_{1}}^{2}+(I_{d_{1}}-1)^{2}}{%
R_{d_{1}}^{2}+(I_{d_{1}}+1)^{2}}\right| _{z \to
z-E_{d_{2}}(k_{2})},  \label{BRI}
\end{equation}
where $R_\nu$ and $I_\nu$ are defined in (\ref{GRI}). This formula
and Lemma \ref{lA1} lead to (\ref{b1}).

\begin{lemma}
\label{lA3} Let $\widehat{b}(k_{2};z)$ be defined by (\ref{bGak}).
Then
\begin{itemize}
\item[(i)]  $\widehat{|b}(k_{2};E+i0)|\leq 1,\ \forall E\in \mathbb{R}$;
\item[(ii)]  for any $\gamma >0$ and $|E|\leq d-\gamma $ there exists an
open set $K_{\gamma }(E)\subset \mathbb{T}^{d_{2}}$, such that
\begin{equation}
\widehat{b}(k_{2};E+i0)<1,\ k_{2}\in K_{\gamma }(E).  \label{pb}
\end{equation}
\end{itemize}
\end{lemma}
\noindent \textit{Proof}. The part (i) of the lemma follows from
Lemma \ref{lA1}, and
from (\ref{BRI}). To prove assertion (ii) we have to find that for any $%
\gamma >0$ and $|E|<d-\gamma $ there exists an open set $K_{\gamma
}(E)$
such that for $k_{2}\in K_{\gamma }(E), \; |E-E_{d_{2}}(k_{2})|<d_1$. Then $%
I_{d_{1}}(E+i0)$ will be strictly positive and
$\widehat{b}(k_{2};E+i0)$ will be strictly less then $1$ in view
of (\ref{BRI}). Since $E_{d_{2}}$ is a continuous function in
$k_2$ on $\mathbb{T}^{d_{2}}$, varying between
$-d_{2}$ and $%
d_{2}$, respective open set $%
K_{\gamma }(E)$ always exists if $|E|<d$. Lemma is proved.

\begin{lemma}
\label{lA4}Let $G_{0 }^{(\nu)}(\x;z)$ be the Green function of the $\nu $%
-dimensional Laplacian and $g>0$. Then the expression
\begin{equation}
\frac{G_{0 }^{(\nu)}(x;E+i0)}{gG_{0 }^{(\nu)}(0;E+i0)+i}
\label{expr}
\end{equation}
is bounded in $x\in \mathbb{Z}^{\nu }$ and in $E\in \mathbb{R}.$
\end{lemma}
\noindent \textit{Proof}. Consider first the one-dimensional case
$\nu =1$. Then it follows from (\ref{G1}) that the expression
(\ref{expr}) is
\begin{equation*}
\frac{e^{i\eta (E+i0)|x|}}{g+\sin \eta (E+i0)},
\end{equation*}
and, according to (\ref{eta}) - (\ref{eta1}), the modulus of the
last expression is bounded by $g^{-1}$.

\medskip \noindent
For $\nu \geq 2$ we will use the integral representation of
$G_{0 }^{(\nu)}(\x;z)$ of (\ref{Go}), valid for $\Im z>0$:
\begin{equation}
G_{0 }^{(\nu)}(x;z)=i\int_0^{\infty} dte^{izt}\prod_{l=1}^{\nu
}J_{x_{l}}(t)e^{i\pi x_{l}/2},  \label{rep}
\end{equation}
where $x=\{x_{l}\}_{l=1}^{\nu }$, and $J_{n}(t)$ is the Bessel
function of the order $n$:
\begin{equation*}
J_{n}(t)=\frac{1}{2\pi }\int_{0}^{2\pi }e^{in\vartheta +it\sin
\vartheta }d\vartheta .
\end{equation*}
The representation follows easily from (\ref{G0}), and from the
identity
\begin{equation}\label{lz}
(\lambda -z)^{-1}=i\int_{0}^{\infty }dte^{-it(\lambda -z)},\
\lambda \in \mathbb{R},\Im z>0.
\end{equation}
By using the asymptotic formula
\begin{equation}
J_{n}(t)=\left( \frac{2}{\pi t}\right) ^{1/2}\cos \left(
t-\frac{(n+1/2)\pi }{2}\right) +O(\frac{1}{t}),\ t\rightarrow
\infty ,  \label{Jas}
\end{equation}
we find that $\nu \geq 3\ G_{0 }^{(\nu)}(x;E+i0)$ is bounded in
$x$ and in $E $. Since, in addition, $|gG_{0 }^{(\nu)}
(0;E+i0)+i|\geq g\Im G_{0}^{(\nu)}(0;E+i0)+1\geq 1$
(recall that in view of (\ref{Go}) $\Im G_{0 }^{(\nu)}(0;z)$ is
nonnegative), we obtain the assertion of the lemma for $\nu \geq
3$.

\medskip \noindent
Thus, we are left with the case $\nu =2$. By using again (\ref{rep}) and (%
\ref{Jas}), we find that $G_{0 }^{(\nu)}(x;E+i0)$ is bounded in
$x$ and in $E $ everywhere except $|E|=2$, and that in a
sufficiently small neighborhood
of $E=2$%
\begin{equation*}
G_{0 }^{(\nu)}(x;E+i0)=A(x)\log |E-2|+B_{\pm }(x)+O(|E-z|), \ E-z
\to 0,
\end{equation*}
where $A(x)$ and $B(x)$ are bounded in $x,$ $A(0)\neq 0,$ and
$B_{\pm }(x)$ correspond to $\mathrm{sign}(E-2) $. The same
asymptotic representation is valid in a neighborhood of $E=-2$.
This shows that the ratio
$G_{0 }^{(\nu)}(x;E+i0)/G_{0 }^{(\nu)}(0;E+i0)$ is bounded and
continuous in $E\in \mathbb{R}$
for any $x\in \mathbb{Z}^{\nu }$. In addition we have:
\begin{eqnarray*}
\left| \frac{G_{0}^{(\nu)}(0;E+i0)}{gG_{0 }^{(\nu)}(0;E+i0)+i}\right|
&=&\left| \frac{1}{g+i\left[ G_{0
}^{(\nu)}(0;E+i0)\right] ^{-1}}\right|  \\ &\leq &\frac{1}{g-
\Im\left[ G_{0 }^{(\nu)}(0;E+i0)\right]
^{-1}}\leq g^{-1},
\end{eqnarray*}
because
\begin{equation*}
-\Im\left[ G_{0 }^{(\nu)}(0;E+i0)\right] ^{-1}
=\Im G_{0 }^{(\nu)}(0;E+i0)/|G_{0 }^{(\nu)}(0;E+i0)|^{2}\geq
0.
\end{equation*}
Lemma is proved.

\begin{lemma}
\label{lA5}The expression
\begin{equation*}
\frac{G_{0}^{(d_{1})}(\x;E-E_{d_{2}}(k_{2})+i0)}{g\gamma_0
(k_{2},E+i0)+i}
\end{equation*}
is bounded in $E\in \mathbb{R},\ k_{2}\in \mathbb{T}^{d_{2}}$, and
$x\in \mathbb{Z}^{d_{1}}$.
\end{lemma}
\noindent
\textit{Proof}. According to (\ref{Go}), $\widehat{\gamma}
(k_{2},z)=G_{0}^{(d_{1})}(0;z-E_{d_{2}}(k_{2}))$. Hence, we can
apply Lemma \ref{lA4}.

%%%%%%%%%%%%%%%%%%%%%%%%%%%%%%%%%%%%%%%%%%%%%%%%%%%%%%%%%%%%%

\bigskip \noindent
\textbf{\large{Acknowledgements.}} L.~P. would like to thank
the Mathematics Department of the University of Wales, Swansea, where a
part of the paper was written.


\begin{thebibliography}{99}
\bibitem{AG} Akhiezer, N.I., Glazman, I.M., ``Theory of Linear Operators in
Hilbert Space,'' F. Ungar, New York, 1963, vol. II.


\bibitem{BS} Boutet de Monvel, A., Surkova, A.,
``Localisation des \'etats de surface pour une classe d'opérateurs
de Schr\"odinger discrets à potentiels de surface
quasi-périodiques,'' 
Helv. Phys. Acta \textbf{71}, 459--490 (1998).


\bibitem{FKS} Cornfeld, I. P., Fomin, S. V., Sinai, Ya. G. 
``Ergodic Theory,'' Springer-Verlag, New York, 1982.

\bibitem{DS} Davies, E. B., Simon, B.,
``Scattering theory for systems with different spatial asymptotics
on the left and right,''  
Comm. Math. Phys. \textbf{63} 277--301 (1978).


\bibitem{PF:84}
Figotin, A. L., Pastur, L. A., 
``An exactly solvable model of a
multidimensional incommensurate structure,''  
Comm. Math. Phys. \textbf{95}, 401--425 (1984).

\bibitem{Gri} Grinshpun, V., 
``Localization for random potentials supported on a subspace,''  
Lett. Math. Phys. \textbf{34}, 103--117 (1995).


\bibitem{Gr}
Grossmann, A., Hoegh-Krohn, R., Mebkhout, M., 
``The one particle theory of periodic point interactions. 
Polymers, mono-molecular layers, and crystals,'' 
Comm. Math. Phys. \textbf{77}, 87--110 (1980).

\bibitem{JL1}
Jak\v si\'c, V., Last, Y., 
``Spectral structure of Anderson type Hamiltonians,'' 
Invent. Math. \textbf{141},  561--577 (2000).


\bibitem{JL2}
Jak\v si\'c, V., Last, Y., 
``Corrugated surfaces and a.c. spectrum,''
Rev. Math. Phys. \textbf{12},  1465--1503 (2000).

\bibitem{JL3} Jak\v si\'c, V., Last, Y.,
``Surface states and spectra,''  
Comm. Math. Phys.  \textbf{218},  459--477 (2001).


\bibitem{JM1}
Jak\v si\'c, V., Molchanov, S., 
``On the spectrum of the surface Maryland model,'' 
Lett. Math. Phys. \textbf{45},  189--193 (1998).


\bibitem{JM2}
Jak\v si\'c, V.,  Molchanov, S., 
On the surface spectrum in dimension two. 
Helv. Phys. Acta \textbf{71}, 629--657 (1998).


\bibitem{JM3}
Jak\v si\'c, V., Molchanov, S., 
``Localization of surface spectra,''
Comm. Math. Phys. \textbf{208},  153--172 (1999).


\bibitem{JM4}
Jak\v si\'c, V., Molchanov, S.,  
``Wave operators for the surface Maryland model,'' 
J. Math. Phys. \textbf{41},  4452--4463 (2000).


\bibitem{JMP}
Jaksic, V.,  Molchanov, S., Pastur, L., 
``On the propagation properties of surface waves,'' 
In: IMA Vol. Math. Appl. \textbf{96}, Springer, New York, 1998, pp. 145-154.


\bibitem{Ka1}
Karpeshina, Yu. E., 
``The spectrum and eigenfunctions of the
Schr\"{o}dinger operator in a three-dimensional space with
point-like potential of the homogeneous two-dimensional lattice
type (Russian),'' 
Teoret. Mat. Fiz.  \textbf{57},  414--423 (1983).

\bibitem{Ka2}
Karpeshina, Yu. E., 
``An eigenfunction expansion theorem for the
Schr\"{o}dinger operator with a homogeneous simple two-dimensional
lattice of potentials of zero radius in a three-dimensional space
(Russian),''  
Vestnik Leningrad. Univ. Mat. Mekh. Astronom. 1984, vyp.1, 11--17.

\bibitem{Ka} Kato, T.,  
``Perturbation Theory for Linear Operators,''
Springer-Verlag, New York, 1966.


\bibitem{KP}
Khoruzhenko, B., Pastur, L., 
``Localisation of surface states: an explicitly solvable model,''  
Physics Reports \textbf{288}, 109-125 (1997).


\bibitem{KST} Kosevich, A. M., Syrkin, E. S., Tutov A. V.,
``Acoustic shear waves localized near a planar defect in an fcc crystal,''
Low Temperature Physics  \textbf{22}, 545-642 (1996).

\bibitem{Lo} Love, A. E. H., 
``A Treatise on the Mathematical Theory of Elasticity,'' 
Dover Publications, New York, 1944.


%\bibitem{Pa}
%Pastur, L, Surface waves: propagation and localisation In:
%\emph{Journ{\'e}es  "Equations aux D{\'e}riv{\'e}es Partielles"},
%(Saint-Jean-de-Monts, 1995), Exp. No. VI, 12 pp.

\bibitem{PF}
Pastur, L., Figotin A.,
``Spectra of Random and Almost Periodic Operators,''
Spriger Verlag, Berlin-Heidelberg, 1992.

\bibitem{Pe}
Pearson, D. B., 
``Quantum Scattering and Spectral Theory,'' 
Academic Press, London, 1988.

 \bibitem{RS}
Reed, M., Simon, B.,  
``Methods of Modern Mathematical Physics. III. Scattering Theory,''  
Academic Press, New York-London, 1979. xv+463 pp.


\bibitem{Si}
Simon, B.,  
``Quantum Mechanics for Hamiltonians Defined as Quadratic Forms,''  
 Princeton University Press, Princeton, N. J..

\bibitem{Si:84}
Simon, B., 
``Almost periodic Schr\"{o}dinger operators. IV. The Maryland model,''  
Ann. Physics \textbf{159},  157--183 (1985).

\end{thebibliography}
\end{document}